\documentclass[final,5p,times,twocolumn,sort&compress]{elsarticle}

\usepackage{amssymb}
\usepackage{dcolumn}
\usepackage{bm}

\usepackage[T1]{fontenc} 

\usepackage{float}
\usepackage{multirow}
\usepackage{amsmath}
\usepackage[dvipsnames]{xcolor}
\usepackage{hyperref}
\hypersetup{
    colorlinks=true,
    allcolors=[rgb]{0.26,0.41,0.88},
}
\usepackage[normalem]{ulem}
\usepackage{xspace}

\newcommand{\gambit}[0]{\textsf{GAMBIT}\xspace}

\journal{Physics Letters B}
\def\tsc#1{\csdef{#1}{\textsc{\lowercase{#1}}\xspace}}
\tsc{WGM}
\tsc{QE}

\begin{document}

\begin{frontmatter}


\title{Electroweak baryogenesis from charged current anomalies in $B$ meson decays}

\author[1]{Peter Athron}
\ead{peter.athron@coepp.org.au}
\author[2,3,4,5]{Michael J.~Ramsey-Musolf}
\ead{mjrm@sjtu.edu.cn, mjrm@physics.umass.edu}
\author[2]{Cristian Sierra\corref{cor1}}
\ead{cristian.sierra@cern.ch}
\author[1]{Yongcheng Wu}
\ead{ycwu@njnu.edu.cn}

\cortext[cor1]{Corresponding author}

\affiliation[1]{organization={Department of Physics and Institute of Theoretical Physics, Nanjing Normal University},
            addressline={Wenyuan Road}, 
            city= {Nanjing},
            postcode={10023}, 
            state={Jiangsu},
            country={China}}

\affiliation[2]{organization={Tsung-Dao Lee Institute \& School of Physics and Astronomy, Shanghai Jiao Tong University},
            addressline={1 Lisuo Road, Pudong New Area}, 
            city= {Shanghai},
            postcode={201210}, 
            country={China}}           
            
\affiliation[3]{organization={Shanghai Key Laboratory for Particle Physics and Cosmology, Key Laboratory for Particle Astrophysics and Cosmology (MOE), Shanghai Jiao Tong University},
            city= {Shanghai},
            postcode={200240}, 
            country={China}} 
            
\affiliation[4]{organization={Amherst Center for Fundamental Interactions, Department of Physics, University of Massachusetts},
            city= {Amherst},
            postcode={01003}, 
            state={MA},
            country={USA}}  
          
\affiliation[5]{organization={Kellogg Radiation Laboratory, California Institute of Technology},
            city= {Pasadena},
            postcode={91125}, 
            state={CA},
            country={USA}}

\begin{abstract}
We demonstrate for the first time that new physics explaining the long standing charged $B$ meson anomalies, $R(D^{(*)})$,  can be the source of CP violation that explains the observed baryon asymmetry of the universe (BAU). We consider the general two Higgs doublet model with complex Yukawa couplings and compute the BAU in the semiclassical formalism, using a novel analytic approximation for the latter. After imposing constraints from both flavor observables and the electron electric dipole moment (eEDM), and assuming the occurrence of a sufficiently strong first order electroweak phase transition, we find that the observed BAU can be generated for benchmark points within the region favored by the global flavor fit. Reaching the observed BAU relies on optimized values of the phase transition parameters, which we treat as inputs. These scenarios, which explain both $R(D^{(*)})$ and the BAU, can be probed  with future eEDM experiments and Higgs factories measurements.
\end{abstract}

\begin{keyword}

Electroweak baryogenesis \sep  Two-Higgs doublet model

\end{keyword}

\end{frontmatter}

\section{Introduction}
\label{introduction}

The origin of the observed baryon asymmetry of the Universe (BAU) is one of the deepest puzzles in particle physics and cosmology. To generate it, the Sakharov conditions \cite{Sakharov:1967dj} must be satisfied: (1) baryon
number violation; (2) C and CP violation and (3) departure from thermal equilibrium. The Standard Model (SM) in principle can fulfill all three conditions, but in practice fails to do so for the latter two. Thus, successful baryogenesis requires physics beyond the SM (BSM).

One of the most attractive mechanisms for this is Electroweak Baryogenesis (EWBG) \cite{Kuzmin:1985mm,Shaposhnikov:1986jp,Shaposhnikov:1987tw,Cline:2006ts,Morrissey:2012db} as it can be tested with electric dipole moments (EDMs)~\cite{Chupp:2017rkp}, collider physics~\cite{Curtin:2014jma,Ramsey-Musolf:2019lsf} and gravitational waves~\cite{Caprini:2015zlo,Huang:2016odd,Beniwal:2017eik}. EWBG requires new bosonic degrees of freedom in order to generate a strong first order phase transition (SFOPT) that breaks electroweak symmetry when the Higgs field develops a non-zero vacuum expectation value (VEV). 
In the SFOPT bubbles of the broken phase nucleate and expand. Non-perturbative electroweak sphaleron processes that change total baryon plus lepton number~\cite{Manton:1983nd,Klinkhamer:1984di} (condition 1) are active outside the bubbles but heavily suppressed inside. The presence of the phase transition and the associated phase boundary (bubble wall) reflect a departure from equilibrium (condition 3). CP violating interactions (condition 2) at the bubble wall generate a chiral asymmetry, i.e., a non zero number density for left handed particles.  This asymmetry generated at the bubble wall subsequently diffuses into the space outside it.  The sphaleron process then converts that chiral asymmetry into a baryon asymmetry that is not washed out when the corresponding particles are swept inside the bubble wall as it expands.   

A BAU compatible with observation can only be predicted if new sources of CP violation are present in the BSM model.  In this work we show that this new CP violation may be sourced from interactions that have been proposed~\cite{Martinez:2018ynq,Crivellin:2023sig,Athron:2024rir} to solve an entirely different problem: long standing and widely considered deviations between SM predictions and the data in flavor physics experiments. Specifically, the $R(D^{(*)})$ ratios in $b\to c\tau\nu$ decays~\cite{HFLAV:2022esi} and the angular observables in $b\to s\mu^{+}\mu^{-}$ transitions~\cite{LHCb:2020lmf,CMS-PAS-BPH-21-002} in $B$ meson decays,  also known as the charged and neutral current anomalies respectively, have garnered considerable attention in flavor physics analyses \cite{Greljo:2022jac,Alguero:2023jeh,Crivellin:2023zui}. Experimental measurements of these observables deviate from SM predictions by roughly $3.8\,\sigma$ and by more than $5\sigma$ when considering global fits to all $b\to s\mu^{+}\mu^{-}$ data~\cite{Greljo:2022jac,Alguero:2023jeh}. Here we investigate the link between the BAU and these persistent signals in the measurements from the LHCb and other particle detectors.

We show for the first time that the charged current anomalies could be revealing new physics that gives rise to a successful EWBG mechanism. This mechanism generates the observed BAU while satisfying  EDM~\cite{Roussy:2022cmp}, flavor and theoretical constraints~\cite{Crivellin:2023sig,Athron:2024rir}.  Several attempts at linking  flavor anomalies to the BAU have been presented in the literature \cite{Liu:2011jh,Tulin:2011wi,Cline:2011mm,Nelson:2019fln,Hou:2023ran}, but none of them have considered the effects in the BAU from new CP-violating couplings entering in the charged current anomalies \cite{Martinez:2018ynq,Crivellin:2023sig,Athron:2024rir}. 
To establish the flavor anomaly-BAU connection, we employ the semi-classical formalism also called Wentzel-Kramers-Brillouin (WKB) method for a leptonic CP-violating source, which provides a more conservative estimate than other approaches \cite{Li:2024mts} as discussed below.

For concreteness, we adopt the general two Higgs doublet model (G2HDM), which includes complex leptonic Yukawas and extra bosonic degrees of freedom. The latter can modify the shape of the Higgs potential away from that of the SM, wherein electroweak symmetry is broken only through a smooth cross-over transition \cite{Kajantie:1996mn}. As a result it has been found both perturbatively \cite{Dorsch:2013wja,Basler:2016obg,Bernon:2017jgv,Dorsch:2017nza,Andersen:2017ika,Kainulainen:2019kyp,Su:2020pjw,Goncalves:2021egx,Biekotter:2022kgf,Goncalves:2023svb} and non-perturbatively \cite{Andersen:2017ika,Kainulainen:2019kyp} that the electroweak phase transition in the 2HDM can be strongly first order.

EWBG in the 2HDM has been studied extensively with both scalar \cite{Cohen:1991iu,Joyce:1994bi,Cline:1995dg,Fromme:2006cm,Fromme:2006wx,Cirigliano:2009yd,Cline:2011mm,Dorsch:2016nrg,Basler:2021kgq,Enomoto:2021dkl,Enomoto:2022rrl,Goncalves:2023svb} and fermion sources \cite{Chung:2008aya,Chung:2009cb,Liu:2011jh,Guo:2016ixx,Chiang:2016vgf,Fuyuto:2017ewj} constrained by flavor physics,
but here 
we compute the BAU from a source that also explains the charged current anomalies. Furthermore, we fit the observed BAU and the flavor anomalies while simultaneously explaining another recent (local) $3\,\sigma$ excess in searches for $t\to b\,H^{+}\to b\bar{b}\,c$ from the ATLAS experiment~\cite{ATLAS:2023bzb} that have been linked to the $B$ meson anomalies in Ref.~\cite{Crivellin:2023sig}. Finally, we show that these scenarios can be probed at future Higgs factories~\cite{CEPCStudyGroup:2018ghi,FCC:2018evy}.

\section{Electroweak baryogenesis}
The dynamics governing the generation of the left-handed fermion number density $n_L$ entails three elements: CP-violating scattering at the bubble wall; particle number density transfer reactions; and diffusion. The corresponding equations that encode these dynamics follow from the general set of Kadanoff-Baym (KB) equations. Solving the full set of KB equations is practically challenging, and various approximations have been taken in the literature. Here, we follow the WKB treatment of Refs.~\cite{Cline:2001rk,Cline:2020jre,Kainulainen:2021oqs} that yields a set of Boltzmann equations which one may further reduce to a tractable diffusion equation for $n_L$.

For the present purposes, a further simplification follows from the work of Ref.~\cite{DeVries:2018aul}. As noted in Ref.~\cite{DeVries:2018aul}, in principle there is a transport equation for each new scalar degree of freedom appearing in the G2HDM. However, the new scalars will reach thermal equilibrium with the SM Higgs as
significant interactions are required for a first order phase transition, and as a result the new particle densities can be added to the Higgs number density. Regarding CP-violation in the model, we will consider it in the lepton Yukawa sector. The associated CP-conserving Yukawa interactions will re-distribute the lepton asymmetry among the left- and right-handed leptons and Higgs particles. Quark Yukawa interactions will then re-distribute the Higgs number density into the left- and right-handed quark number densities. Strong sphaleron processes will lead to further rebalancing between the left- and right-handed quark densities. Thus, to obtain the resultant left-handed number density relevant to weak sphaleron processes, one must solve a set of coupled Boltzmann equations for the Higgs density and left- and right-handed lepton and quark densities. For a detailed discussion in the case of lepton-mediated EWBG, see Ref.~\cite{Chung:2009cb}. In practice, for this case the authors of Ref.~\cite{DeVries:2018aul} find that $n_L(x)$ can be dominated by the density of left-handed leptons, $l_L(x)$, so that to an order ten percent level of accuracy, one may consider a single Boltzmann equation for $l_L(x)$. Following the prescription of Refs.~\cite{Cline:2001rk,Cline:2020jre,Kainulainen:2021oqs} in the WKB formalism and using the diffusion approximation, one then obtains the single species diffusion equation

\begin{equation}
\overline{D}_l\,l''(z)+\upsilon_{w}\,l'(z)+\overline{\Gamma}_l\,l(z)=\overline{S}_l,\label{eq:WKB_diffusion}
\end{equation}
where we have dropped the subscript \lq\lq $L$\rq\rq\, for simplicity. Here, $\upsilon_{w}$ denotes the bubble wall velocity and $\overline{D}_l$ the effective diffusion constant which is proportional to the diffusion coefficient $D_l$ (expressing how the particles spread out), with modifications to account for the effect of non-zero $\upsilon_{w}$;  $\overline{\Gamma}_l$ is the modified collision rate from inelastic scattering with the wall; and $\overline{S}_l$ is the CP-violating source term. Within the WKB method, the latter arises at second order in derivatives. On the other hand, in the VEV resummation (VR) framework~\cite{Cirigliano:2009yt,Cirigliano:2011di, Li:2024mts}, the source term is first order in derivatives, so we expect our approach to result in conservative (under-)estimates of the BAU. For the explicit expressions for all the terms in Eq.(\ref{eq:WKB_diffusion}) the reader is referred to the appendix, Eq.(\ref{eq:coefficients}). 

We now comment further on the assumptions leading to the aforementioned simplified treatment.
First, the lepton Yukawa interactions are sufficiently slow that the transfer of lepton number densities into other species is relatively inefficient~\cite{DeVries:2018aul}.
The diffusion equations for both right-handed and left-handed leptons can be simplified into a single equation that decouples from both quarks and the Higgs boson, the latter because of the slow lepton Yukawa rate $\Gamma_{Y_{\tau}}$ as shown in Ref.~\cite{DeVries:2018aul}. 
Second, we take the left- and right-handed diffusion constants to be approximately equal: $D_l\equiv D_{L}=D_{R}\simeq100/T$, which implies $l\equiv l_L=-\tau_R$.
This was proven to be valid up to a $10\%$ difference with respect to
the exact treatment of having different diffusion constants, both analytically and numerically~\cite{DeVries:2018aul}. Given that $\overline{\Gamma}\gg\upsilon_{w}^{2}/D_{l}$, Eq.(\ref{eq:WKB_diffusion}) can
be solved analytically using the Green's function method. We neglect the bubble wall curvature and work in the bubble wall rest frame, $z=|\boldsymbol{x}+\boldsymbol{\upsilon}_w\,t|$ and we take $z<-2L_{w}$ as the symmetric phase, where $L_{w}$ is the bubble wall thickness. In this way, the solution to Eq.(\ref{eq:WKB_diffusion}) is given by~\cite{Huet:1995sh,Lee:2004we,Cirigliano:2006wh,Chung:2009cb,DeVries:2018aul,Xie:2020wzn}
\begin{equation}
l(z)=-\mathcal{A}\,e^{z\,\gamma_{+}^{s}},
\label{eq:lepton_density}\end{equation}
with the amplitude $\mathcal{A}$ given in Eq.(\ref{eq:Amplitude}).

The net baryon number density $n_B$ is determined by solving the baryon number diffusion equation~\cite{Cohen:1994ss},
\begin{equation}
n_B''(z)-\frac{\upsilon_w}{D_q}n_B'(z)=\frac{\Gamma_{ws}}{D_q}\left( \mathcal{R}\,n_B(z)+\frac{3}{2}n_L(z)\right),\label{eq:baryon_diffusion}
\end{equation}
with $D_{q}=7.2/T$~\cite{Cline:2021dkf} the quark diffusion
constant, $\Gamma_{ws}=6.3\times 10^{-6}\,T$ the electroweak sphaleron rate~\cite{Arnold:2000dr} and $\mathcal{R}=15/4$ is the SM relaxation term. Here, the diffusion of the baryon density is sourced by the left-handed leptons $n_L(z)\simeq l(z)$ from Eq.(\ref{eq:lepton_density}) and the solution of Eq.~(\ref{eq:baryon_diffusion}) is given in terms of the ratio $Y_{B}\equiv n_B/s$~\cite{Cohen:1994ss},
\begin{equation}
Y_{B}=-\frac{3\,\Gamma_{ws}}{2\,s}\frac{1}{D_{q}\lambda_{+}}\int_{-\infty}^{-2L_{w}}\,l(z)\,e^{-\lambda_{-}z}dz\label{eq:BAU},
\end{equation}
where $s=(2\pi^{2}/45)g_{*}T^{3}$ is the entropy density at the temperature $T$, $g_{*}=110.75$ are the degrees of freedom in the G2HDM and $\lambda_{\pm}=1/2D_{q}\,(v_{w}\pm\sqrt{v_{w}^{2}+4\,\mathcal{R}\,\Gamma_{ws}\,D_{q}})$. In the previous derivation, we have followed Refs.~\cite{Liu:2011jh,Guo:2016ixx} and worked in the mass basis defined by Eqs.(\ref{eq:mass-fermions},\ref{eq:YukawaLagrangian}) and Eq.(\ref{eq:Utransformation}) in the appendix, with the CP phases only in the lepton sector for massless neutrinos and therefore no new topological terms are induced in the transport equations.

\section{Hints from flavor Physics}
Lepton flavor universality (LFU) is a fundamental
feature of the SM and states that the $W$ boson couples equally to the three lepton generations. It is tested experimentally by considering ratios of semileptonic $B$ meson decays to $\tau$ and light leptons $l$ (either $e$ or $\mu$) defined as
\begin{equation}
    R(D^{(*)}) = \frac{\mathrm{BR}(\bar{B}\to D^{(*)}\tau\bar{\nu})}{\mathrm{BR}(\bar{B}\to D^{(*)}l\bar{\nu})},
\end{equation}
with the latest experimental world average given by $R(D) = 0.347 \pm 0.025, \quad R(D^*) = 0.288 \pm 0.012$~\cite{HFLAV2024winter}. This implies a deviation of $3.8\,\sigma$ from the SM, thus motivating BSM physics in order to fit it. These ratios can be explained at tree level using complex lepton Yukawa couplings in the G2HDM while simultaneously satisfying all other flavor constraints at the $1\sigma$ level~\cite{Martinez:2018ynq,Crivellin:2023sig,Athron:2024rir}. Explicitly, using the form factors provided by \textsf{SuperIso 4.1}~\cite{deDivitiis:2007ptj,Kamenik:2008tj,Mahmoudi:2007vz,Mahmoudi:2008tp,Mahmoudi:2009zz}, in the G2HDM they are,
\begin{equation}
R(D)=\frac{1+1.73\,\mathrm{Re}(g_{S}^{\tau\tau})+1.35\sum\left|g_{S}^{\ell\tau}\right|^{2}}{3.27+ 0.57\,\mathrm{Re}(g_{S}^{\mu\mu})+4.8\sum|g_{S}^{\ell\mu}|^{2}},\label{eq:RD}
\end{equation}
\begin{equation}
R(D^{*})=\frac{1+0.11\,\mathrm{Re}(g_{P}^{\tau \tau})+0.04\sum\left|g_{P}^{\ell\tau}\right|^{2}}{4.04+0.08\,\mathrm{Re}(g_{P}^{\mu\mu})+0.25\sum|g_{P}^{\ell\mu}|^{2}}\label{eq:RDstar},
\end{equation}
with $\ell=e,\,\mu,\,\tau$ and the scalar and pseudoscalar couplings $g_{S,P}^{\ell\ell'}$ given at tree level in Refs.~\cite{Athron:2021auq,Athron:2024rir}. The latter are functions of the Yukawa textures with the complex phase $\theta$ defined in Eq.(\ref{eq:lepton_mass_texture}) in the appendix. 

The primary impact of the neutral current anomalies is to constrain the quark sector, given the dominance of the extensively-studied top quark penguin contributions. They contribute to the vast set of observables entering the global fit, but otherwise have no role in the BAU.

\begin{figure}[h]
\begin{centering}
\includegraphics[scale=0.15]{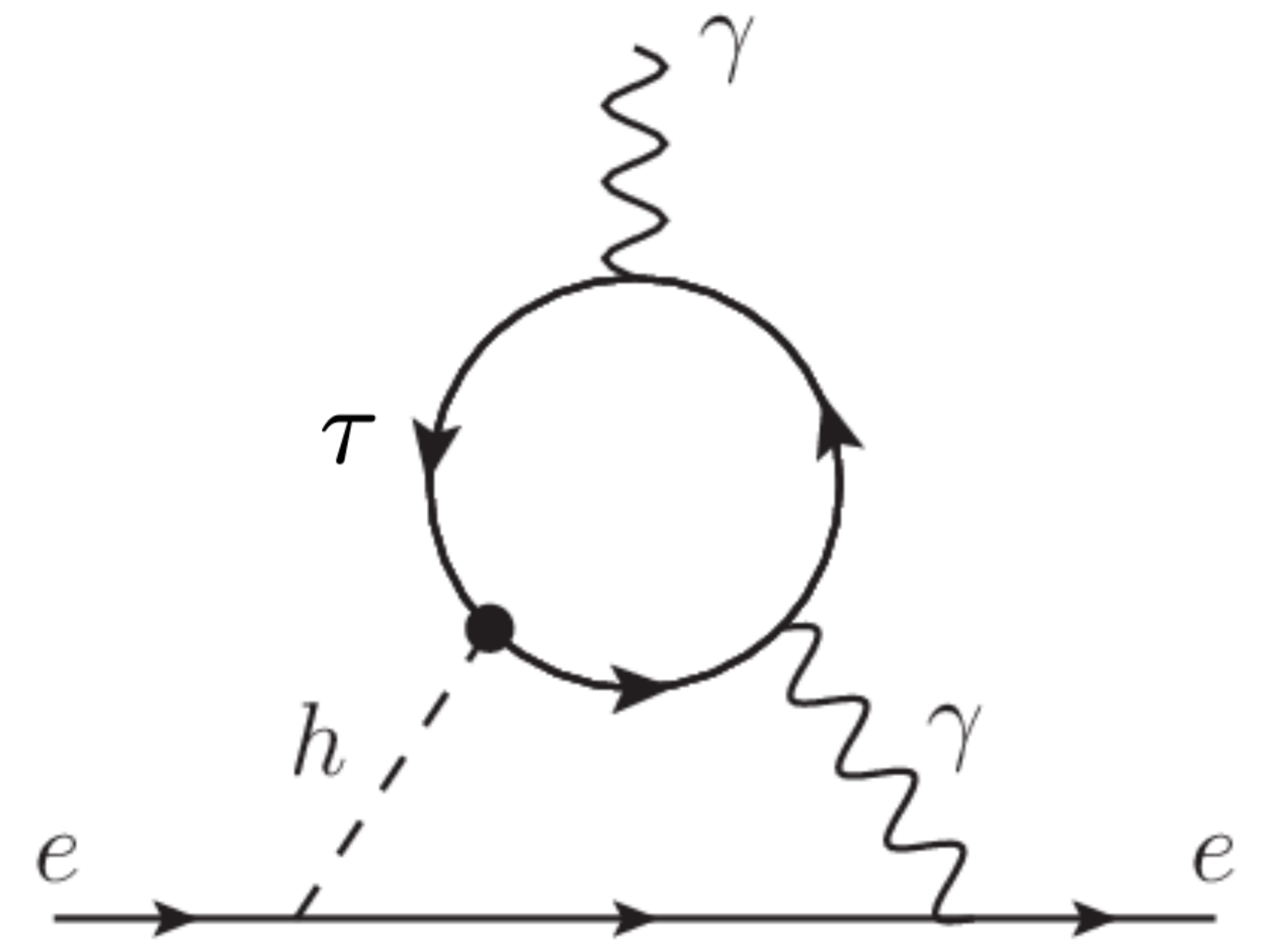}
\par\end{centering}
\caption{\emph{Dominant two-loop Barr-Zee diagram with a $\tau$ lepton in the loop entering in the eEDM calculation.}\label{fig:Barr-Zee}}
\end{figure}

We explore the possibility that complex phase $\theta$ can be large enough to explain the BAU, with the real and imaginary parts of the Yukawa couplings in Eq.~(\ref{eq:lepton_mass_texture}) still accommodating the $R(D^{(*)})$ results and EDM constraints. Regarding the latter, the Higgs fermion effective Lagrangian can be written as~\cite{Alonso-Gonzalez:2021jsa,Ge:2020mcl},
\begin{equation}
-\mathcal{L}_{h\tau\tau}=\frac{m_{\tau}}{v}\left(\kappa_{\tau}\overline{\tau}\tau+i\tilde{\kappa}_{\tau}\overline{\tau}\gamma_{5}\tau\right)h,
\end{equation}
allowing a matching between this effective theory and the G2HDM (see appendix, Eq.(\ref{eq:YukawaLagrangian}) for the relevant part of the G2HDM Lagrangian), obtaining
\begin{align}
\kappa_{\tau}&=-\frac{\sin\alpha}{\cos\beta}+\frac{c_{\beta-\alpha}}{\cos\beta}\,{\rm Re}(N_{\tau\tau}),\nonumber\\
\tilde{\kappa}_{\tau}&=\frac{c_{\beta-\alpha}}{\cos\beta}\,{\rm Im}(N_{\tau\tau}),
\label{Eq:kappas}
\end{align}
where the CP-violating coupling $\tilde{\kappa}_{\tau}$ is given in terms of 
${\rm Im}(N_{\tau\tau})={\rm Im}(J_{A})/y_{\tau}^2$, and $J_{A}$ is the Jarlskog invariant defined in Eq.(\ref{eq:Jarlskog_invariant}) in the appendix. The CP-conserving coupling
$\kappa_{\tau}$ governs the Higgs signal strength that is relevant for future Higgs factories.  It is written here in terms of ${\rm Re}(N_{\tau\tau})=(v/\sqrt{2}\,m_{\tau})Y^{\tau\tau}_l$, with  $Y^{\tau\tau}_l$ given in Eq.(\ref{eq:Gammafphiba}).

Note that $\tilde{\kappa}_{\tau}$ directly contributes to both the electron EDM $|d_e|$ and also to the light quark EDMs through the two-loop Barr-Zee diagram in Fig.~\ref{fig:Barr-Zee}, which is proportional to the corresponding fermion mass ($m_e$ or $m_q$). The light quark EDMs can further contribute to the neutron EDM (nEDM) with $\mathcal{O}(1)$ coefficient~\cite{Hisano:2012cc,Hisano:2012sc,Engel:2013lsa,Bertolini:2019out}. Therefore the contribution of the CP-violating coupling $\tilde{\kappa}_\tau$ to both the eEDM and nEDM could be comparable at first. However, the latest experimental measurements give $|d_e|<4.1\times10^{-30}\,e{\rm cm}$~\cite{Roussy:2022cmp} and $|d_n|<1.8\times10^{-26}\,e{\rm cm}$~\cite{Abel:2020pzs} at 90\% C.L., which differ by about 4 orders of magnitude. Hence, the eEDM is the dominant constraint in our scenario. 

Comparing to the results of Ref.~\cite{Fuchs:2020uoc} for a $\tau$ lepton in the upper loop of the two-loop Barr-Zee diagram, the latest experimental measurement of the eEDM provides $|\tilde{\kappa}_\tau|\lesssim 0.078$. Regarding the future sensitivity on the eEDM, the next generation ACME III experiment will improve it by another order of magnitude to $\delta d_e\approx 4\times 10^{-31}\,e{\rm cm}$~\cite{Ang:2023uoe}, which provides then $|\tilde{\kappa}_\tau|\lesssim 0.0078$. In addition, the lepton Yukawa couplings will be further constrained by other $B$ decays related to the neutral current anomalies, as well as from kinetically allowed Higgs flavor violating decays like $h\to \ell\,\ell'$ and lepton flavor violating decays of the form $\ell\to \ell'\gamma$ for $\ell,\,\ell'=e,\,\mu,\,\tau$~\cite{Athron:2021auq,Kanemura:2023juv,Crivellin:2023sig,Athron:2024rir}.

\section{Results}
We now demonstrate that the G2HDM Yukawas shown by global fits to explain the $R(D^{(*)})$ anomalies and all other relevant constraints, can source sufficient CP violation to also explain the BAU. We find solutions that can do this while satisfying all the flavor constraints discussed above, as well as unitarity, stability and perturbativity bounds and constraints from electroweak precision observables via the $S$, $T$ and $U$ parameters~\cite{ParticleDataGroup:2024cfk}.  
The global fit was obtained by running \gambit~\cite{Athron:2017ard,Martinez:2017lzg,GAMBITFlavourWorkgroup:2017dbx,GAMBITModelsWorkgroup:2017ilg,GAMBIT:2017qxg} with \textsf{SuperIso 4.1}~\cite{Mahmoudi:2007vz,Mahmoudi:2008tp,Mahmoudi:2009zz}, \textsf{2HDMC 1.8}~\cite{Eriksson:2009ws} and \textsf{HEPLike}~\cite{Bhom:2020bfe}.  
The scan follows the analysis of Ref.\ \cite{Athron:2024rir}, specifically the scans around BM3 (originally defined in Ref.\ \cite{Crivellin:2023sig}) with all the likelihoods included there plus a new likelihood introduced here for the eEDM from the JILA-NIST group bound on $|d_{e}|$~\cite{Roussy:2022cmp}.  
In general, the observables with the smallest uncertainties and those associated with the largest set of observables included in the experimental measurements drive the global fits.

\begin{figure}[h]
\begin{centering}
\includegraphics[scale=0.6]{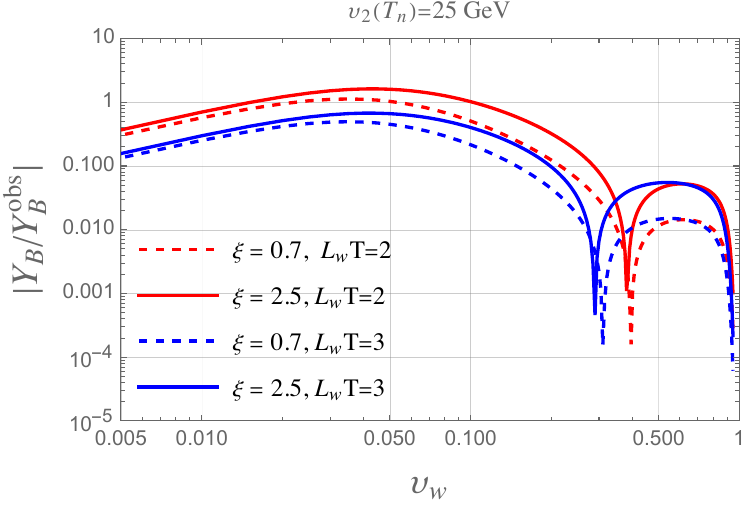}
\par\end{centering}
\caption{\emph{Ratio $|Y_{B}/Y_{B}^{{\rm obs}}|$ as a function of $\upsilon_w$ with fixed values of the extra Yukawa couplings within the 95 $\%$C.L region of the BM3 study of \cite{Athron:2024rir}, $y_{\mu\mu}=0.001$,
$y_{\mu\tau}=0.01$, $y_{\tau\tau}=0.06$, $\theta=2.1$ and $c_{\beta-\alpha}=0.01$.}\label{fig:BAU_vw}}
\end{figure}

To determine the BAU from Eq.(\ref{eq:BAU}), we need to know the values of the so-called thermal parameters, quantities that characterize the FOPT. These thermal parameters can be determined from the fundamental G2HDM parameters in the Higgs potential, see Refs.\ \cite{Goncalves:2021egx,Biekotter:2022kgf,Goncalves:2023svb} for recent studies. We simply select reasonable choices for the thermal parameters based both on previous work \cite{Goncalves:2021egx,Biekotter:2022kgf,Goncalves:2023svb} and an understanding of how they influence the BAU calculation. For clarity we collect here the assumptions and unknowns of our BAU computation. First, a strong first order electroweak phase transition with $\xi\equiv v_n/T_n$ of order unity is assumed but not derived from the G2HDM potential. Second, the bubble wall width $L_w$ and velocity $\upsilon_w$ are treated as independent thermal parameters, and reaching the observed BAU requires thin walls ($L_wT_n\simeq2$--$3$) and slow walls ($\upsilon_w\lesssim0.1$). Third, the CP-violating phase $\theta$ is taken at its best fit value, $|\theta|\simeq2.1$, so the BAU shown corresponds to optimized points within the flavor-favored region. Finally, our WKB estimate is conservative with respect to the VEV-resummed framework. Identifying the scalar potential parameters that realize these thermal parameters is left to future work.

For all scenarios we fix $v_2(T_n) =25\,\mathrm{GeV}$ which is a small amplitude developed by the field profile of the second Higgs only within the bubble wall (see Eq.\ (\ref{eq:kink})) consistent with values found in \cite{Kanemura:2023juv}. We obtain the source term on the right hand side of Eq.\ (\ref{eq:baryon_diffusion}) using the expressions given in the appendix (see in particular Eqs.\ (\ref{eq:coefficients}) and (\ref{eq:tau_source})). We then compute the analytical chiral lepton-antilepton asymmetry in Eq.~(\ref{eq:lepton_density}) and proceed to numerically determine the BAU for different choices of the bubble wall velocity expressed as a fraction of the speed of light, $\upsilon_w$, the bubble wall thickness, $L_w$, and the ratio $\xi\equiv v_n/T_n$, where $T_n$ is the nucleation temperature and $v_n=v_1(T_n)$ is the electroweak VEV at $T_n$, and we simply fix $v_n = 200$ GeV since our results are only sensitive to the ratio $\xi$. Regarding the latter, Ref.~\cite{Goncalves:2021egx} found that $1<\xi<3.5$ is allowed in 2HDMs. It is worth mentioning here that the conventional criteria  $v_n/T_n >1$ is (a) theoretically problematic, as it is not in general gauge invariant and (b) the real issue is preservation of the baryon asymmetry produced at the onset of the electroweak phase transition. In both respects, this conventional criterion has significant uncertainties, and a more rigorous treatment as in Ref. \cite{Patel:2011th} or the more recent EFT formulation in Ref. \cite{Li:2025kyo} should be considered. That said, for our present purposes we consider $v_n/T_n$ of order unity.

First we consider the BAU that can be obtained with the Yukawa textures at the best fit point from the global fit (see Eq.(\ref{eq:lepton_mass_texture}) for details). Fig.~\ref{fig:BAU_vw} shows the BAU (specifically $Y_{B}/Y_{B}^{{\rm obs}}$ with $Y_{B}^{{\rm obs}}=8.59\times10^{-11}$ being the central value given by the Planck Collaboration~\cite{Planck:2013pxb}) plotted against the bubble wall velocity for different values of parameters characterizing the phase transition. The dashed (solid) curves are plotted for $\xi=0.7\,(2.5)$, with the red versions for $L_w T_n=2$, and the blue versions for $L_w T_n=3$.
These show that successful EWBG is achieved for low $\upsilon_w$ for a sufficiently SFOPT $\xi\gtrsim 0.7$ and relatively thin bubble wall widths $L_w T_n\gtrsim 2$, where the WKB approximation is still valid. More precisely, the WKB approach is an expansion in $(L_wT_n)^{-1}$ and requires $1\ll L_wT_n$ for the wall to be thick compared to the typical de Broglie wavelength $\sim T^{-1}$ of the plasma particles~\cite{Cline:2001rk,Cline:2020jre,Basler:2021kgq,Enomoto:2022rrl}, and our choice $L_wT_n=2$--$3$ therefore lies within this validity range, with higher-order corrections at most ${\cal O}(1)$. Furthermore, dedicated wall-profile computations in the C2HDM find larger widths for light Higgs spectra, $L_wT_c\simeq4$--$115$~\cite{Basler:2021kgq}. Since the source scales roughly as $1/L_w^2$, as displayed by the $L_wT_n=3$ curves in Fig.~\ref{fig:BAU_vw}, our thin-wall choice is on the optimistic side and should be regarded as one of the optimized assumptions of our analysis. The sharp dip in the BAU near $\upsilon_w=0.3$ in Fig.~\ref{fig:BAU_vw} is due to cancellations between terms with different signs in the WKB calculation~\cite{Cline:2020jre,Cline:2021dkf}. 

Our results at first contrast with the values obtained in Ref.\cite{Cline:2021dkf}, which for a singlet model found the baryon abundance for low $\upsilon_w$ to be of order $\sim4\times10^{-3}$. We corroborated that for similar values of the parameters used there, our calculations gave the same order of magnitude. The key differences between the singlet and the two Higgs doublet calculations rely on first, having larger extra Yukawa couplings and second, a different scaling in the WKB source in the singlet model proportional to $1/\Lambda$, which itself suppressed the baryon abundance by a factor of $10^3$. In addition to this, we are also using thinner bubble walls compared to them and a stronger transition strength, which combined with the previous effects give a net increase in the G2HDM with respect to the singlet model of a factor $\sim2\times10^3$.

\begin{figure}[h]
\begin{centering}
\includegraphics[scale=0.23]{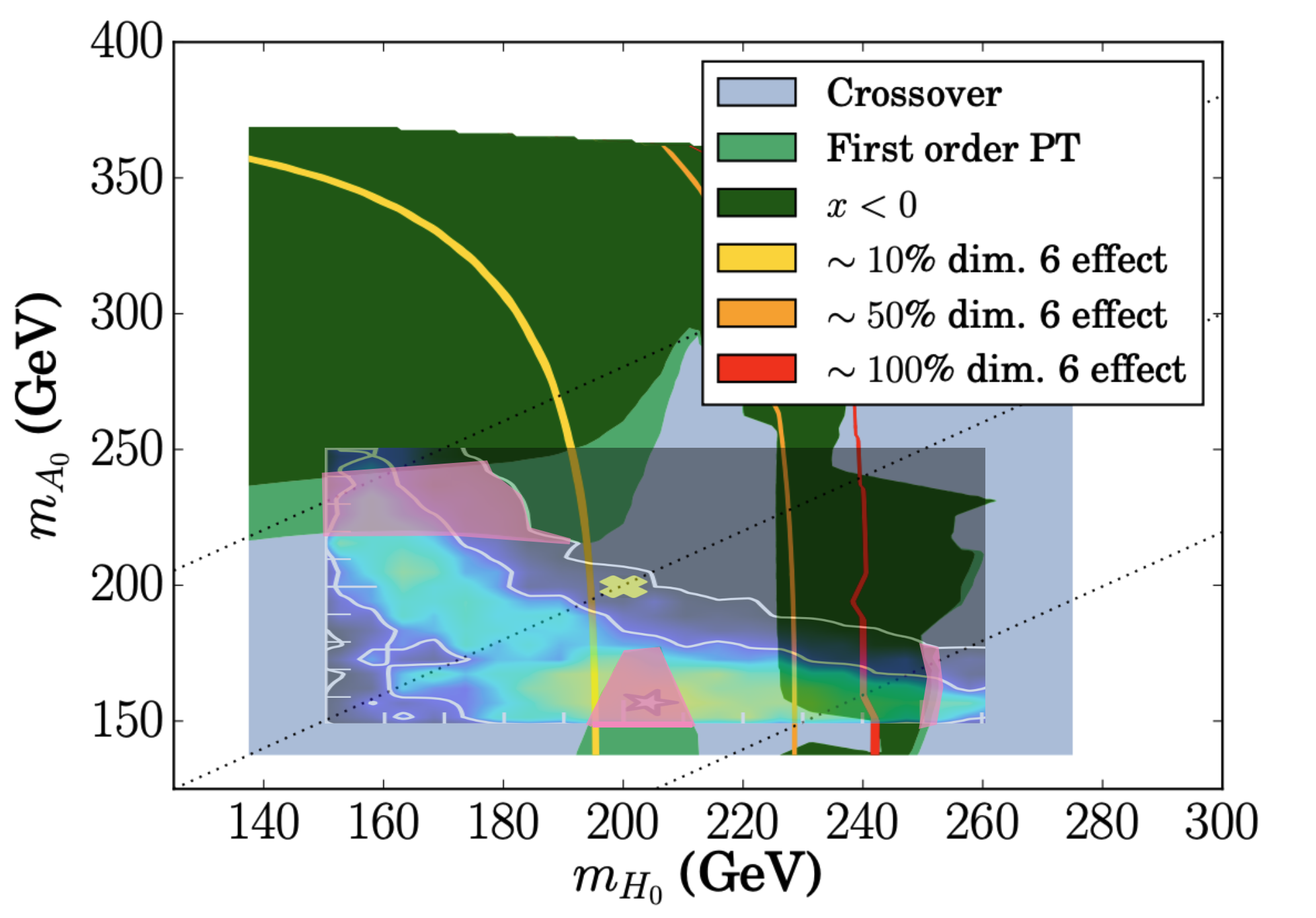}
\par\end{centering}
\caption{\emph{Overlapped $m_H-m_A$ plane from Figure 2 and Figure 3 of \cite{Andersen:2017ika} and \cite{Athron:2024rir} respectively. The pink regions show the common parameter space between the plots where a first order phase transition is possible.}\label{fig:overlap}}
\end{figure}

We observe that the parameter space constrained by the global fit is indeed compatible with a SFOPT. The plane of interest from our global fit is the $m_H-m_A$ plane, shown in the left-middle panel of Fig. 3 in Ref.~\cite{Athron:2024rir}. When projecting it onto the respective $m_H-m_A$ plane of Figure 2 of \cite{Andersen:2017ika} corresponding to $\xi=0.7$ there, we obtain overlapping regions in the ranges $m_A\in[210,\,250]$ GeV and $m_H\in[190,\,260]$ GeV which include our best fit point. We show this explicitly in Figure \ref{fig:overlap} as the pink regions when overlapping both figures. We can see that our best fit point (star) is well inside the allowed values for a first order phase transition. The yellow cross is a reference value at (200 GeV, 200 GeV) from \cite{Crivellin:2023sig} that corresponds to the benchmark point there called BM3. 

It is worth noting that it is possible to increase the mass splittings to raise  $\xi$ while explaining the charged current anomalies, as has been shown in Ref.\ \cite{Athron:2021jyv}. However, the global fit in Ref. \cite{Athron:2024rir} was fitting additional ATLAS $t\to H^\pm b$ decay data \cite{ATLAS:2023bzb} which required a light charged Higgs boson at around 130 GeV. Without this constraint, large splittings are still possible.

\begin{figure}[h]
\begin{centering}
\includegraphics[scale=0.65]{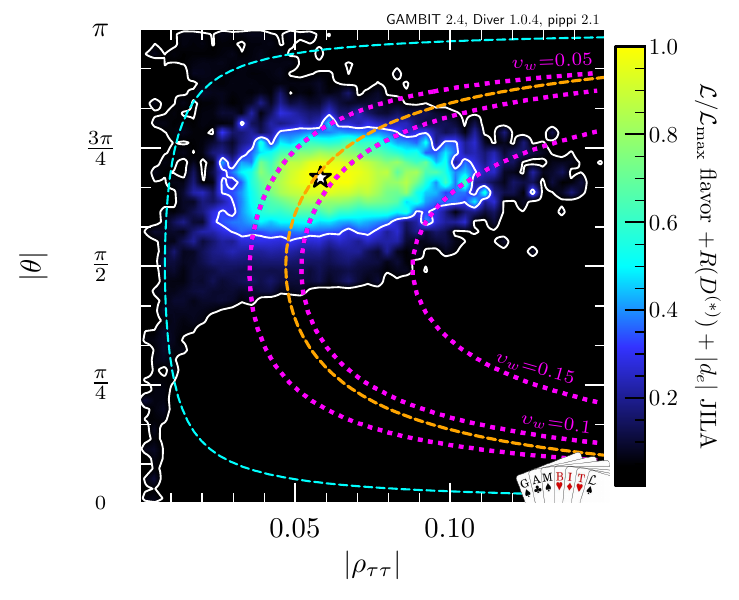}
\par\end{centering}
\caption{\emph{Profile likelihood ratios $\mathcal{L}/\mathcal{L}_{\mathrm{max}}$ from all flavor constraints in \cite{Crivellin:2023sig} and the JILA-NIST bound on $|d_{e}|$~\cite{Roussy:2022cmp}. The white star denotes the best fit point and white contours around it are the $1\sigma$ and $2\sigma$ confidence intervals.  Overlayed on top of this are magenta horizontal dashed lines showing results for the BAU (using best fit values of the parameters from the global fit) for different values of $\upsilon_w$.  These correspond to $Y_{B}/Y_{B}^{{\rm obs}}=1$ with $\xi= 2.5$ and $L_w T_n=2$. The orange dashed line is the sensitivity projection for $|\tilde{\kappa}_{\tau}/\kappa_{\tau}|$ related to the CP-violating Yukawa coupling measurement  
from both the CEPC and the FCC-ee. The cyan dashed curve is the future sensitivity on the eEDM at the ACME III experiment~\cite{Ang:2023uoe}}.}
\label{fig:BAU_constraints}
\end{figure}

In Fig.~\ref{fig:BAU_constraints} we plot global fitting results that show solutions for the BAU coinciding with the allowed region that explains the flavor anomalies in the $|\rho_{\tau\tau}|-|\theta|$ plane of the G2HDM. 
The function $\mathcal{L}$ on the color bar is the sum of all flavor and eEDM likelihoods defined in \gambit and is normalized with respect to the maximum likelihood obtained over the parameter space of the G2HDM. The mass of the charged Higgs boson $m_{H^{\pm}}$ mediating the tree level exchange $b\to\,c\tau\nu_{l}$ has been fitted close to $130\,\mathrm{GeV}$, matching the local excess in searches for $t\to\,b\,H^{+}\to\,b\bar{b}\,c$ with a local significance of $3\,\sigma$ as reported by the ATLAS experiment~\cite{ATLAS:2023bzb}. Furthermore, regarding constraints from the $B_c$ meson lifetime, the fit gives $\mathrm{BR}(B_c\to\tau\bar\nu)\leq60\%$ for all sampling points and $30\%$ at the best fit point value, which is within the current theoretical estimates according to Ref.~\cite{Blanke:2018yud}.

In this way, fixing $\xi= 2.5$ and $L_w T_n=2$ and all Yukawa couplings to the best fit value obtained in the global fit, we draw in Fig.~\ref{fig:BAU_constraints} contours of $Y_{B}/Y_{B}^{{\rm obs}}=1$ for different bubble wall velocities (dashed magenta lines).  We find that the contours of constant $\upsilon_{w}$ lie on an ellipsoid-like surface with the largest contour occurring for $\upsilon_{w}\simeq0.05$. Contours of smaller and larger $\upsilon_{w}$ would overlap in the plot, so we only plot those for larger values of $\upsilon_{w}$. Successful EWBG can be achieved while satisfying all flavor constraints and the best fit value predicts $\upsilon_{w}\approx0.1$ for $|\theta|\simeq 2.1$, with velocities $\upsilon_{w}>0.15$ being disfavored for those fixed values of the Yukawa couplings, the $v_n/T_n$ ratio and the perturbation parameter  $L_w T_n$. We stress that within our conservative WKB estimate the observed BAU requires slow walls, $\upsilon_w\lesssim0.1$. Such velocities are motivated for strong transitions with sizable Higgs mass splittings, where the friction on the wall is enhanced and deflagration solutions are generic~\cite{Dorsch:2021nja}, and $\upsilon_w=0.1$ is adopted as a benchmark value in dedicated C2HDM studies~\cite{Basler:2021kgq}. A first-principles computation of $\upsilon_w$ for our benchmark points is left to future work.

Finally we note that while we are confident that our use of the WKB approach conservatively demonstrates we can obtain a sufficiently large BAU while satisfying EDM limits, the fact that it gives a conservative (under-)estimate of the BAU, in comparison to the VR approach, means that a different BAU calculation may allow for higher bubble wall velocities.  Thus care should be taken in inferring any specific bounds from our plots.
 
Interestingly, sensitivity projections from the future Higgs factories, CEPC~\cite{CEPCStudyGroup:2018ghi} and FCC-ee~\cite{FCC:2018evy} (orange dashed curve), indicate that future experimental studies of Higgs di-tau decay properties sensitive to $\tilde{\kappa}_{\tau}$ can test our predictions.  Specifically, we assume 5 ab$^{-1}$ integrated luminosity to obtain a projected sensitivity to $|\tilde{\kappa}_{\tau}/\kappa_\tau|\lesssim 0.05$~\cite{Ge:2020mcl,Chen:2017bff,Chen:2017nxp,Jeans:2018anq,Altakach:2022ywa,Gritsan:2022php,ILCInternationalDevelopmentTeam:2022izu} as indicated by the orange dashed curve in Fig.~\ref{fig:BAU_constraints}. Furthermore, the future sensitivity on the eEDM at the ACME III experiment~\cite{Ang:2023uoe} (cyan dashed curve) indicates it could probe almost all of the parameter space of our current scenario.

\section{Outlook}
Our study of EWBG in the G2HDM provides a proof of principle that new physics explaining the $R(D^{(*)})$ anomalies can admit sufficient CP-violation as needed to explain the BAU while respecting eEDM and other phenomenological constraints. To our knowledge, this work represents the first time such a connection between the charged current anomalies and successful EWBG has been demonstrated. Our finding relies on using the WKB method for a conservative estimate of the BAU, while employing a new approach that gives an analytical solution through simplifying approximations. This work can be generalized for more complicated cases using different lepton diffusion constants $D_L\neq D_R$, as well as including quarks in the full transport equations that were updated in Ref.~\cite{Cline:2020jre}. 
More complete numerical studies of the G2HDM, including a computation of the BAU in the VR framework, identifying the values of the scalar potential parameters needed for a SFOPT, and determining their impact on the bubble wall profile and velocity, would yield a more quantitatively robust connection between the BAU and flavor anomalies.

\section*{Acknowledgements}
We thank B. Laurent for useful correspondence. We acknowledge the EuroHPC Joint Undertaking for awarding this project access to the EuroHPC supercomputer LUMI, hosted by CSC (Finland) and the LUMI consortium through a EuroHPC Extreme Scale Access call. The work of PA is supported by the National Natural Science Foundation of China (NNSFC) under grant No.\ 12150610460, the NNSFC Key Projects grant No.\ 12335005 and the supporting fund for foreign experts grant wgxz2022021. YW is supported by NNSFC under grant No.\ 12305112. The work of  MJRM was supported by the National Natural Science Foundation of China (NNSFC) grant numbers 12375094 and W2441004. The work of CS was supported by the NNSFC grant number W2633003 and the
Shanghai Post-doctoral Excellence Program grant No. 2025384.

\appendix

\section{Diffusion equation}
The coefficients and source term of the diffusion equation in Eq.(\ref{eq:WKB_diffusion}) are given by
\begin{align}
\overline{D}_l & =\frac{-D_{2}+D_{0}\,\upsilon_{w}^{2}}{\Gamma_{\mathrm{tot},\,l}\,D_{0}},\nonumber \\
\overline{\Gamma}_l & =k_{l}\frac{K_{0}}{D_{0}}\Gamma_{\mathrm{eff}},\nonumber \\
\overline{S}_l & =-k_{l}\frac{T^{2}}{6}\left(\frac{S_{1}}{D_{0}}+\frac{\upsilon_{w}S_{1}'-S_{2}'}{\Gamma_{\mathrm{tot},\,l}\,D_{0}}\right),
\label{eq:coefficients}\end{align}
where $K_0\simeq1.1$ for all bubble wall velocities and temperatures in the massless case. As shown in Ref.~\cite{Lee:2004we}, the statistical $k_l$ factors are in general functions of the thermal masses which for chiral leptons are well approximated by the massless case, i.e., $k_L(0)=2$ and $k_R(0)=1$ obtained when assuming local thermal equilibrium and expanding the number densities $l$ in small chemical potentials $\mu_{l}=(1/k_{l})(6/T^{2})\,l$.

The general expressions for the source functions $S_{1,2}$ in Eq.(\ref{eq:coefficients}) are given in Ref.~\cite{Cline:2020jre} and are model-dependent. The total interaction rate $\Gamma_{\mathrm{tot},\,l}$ is defined as 
\begin{equation}
\Gamma_{\mathrm{tot},\,l}=\frac{D_2(\upsilon_w)}{D_0\,D_l},
\end{equation}
with $D_0=1$ and 
\begin{equation}
D_2(\upsilon_w)=\frac{\upsilon_w\left(2 \upsilon_{w}^2-1\right)+\left(\upsilon_w^2-1\right)^2 \tanh ^{-1}\upsilon_w}{\upsilon_w^3},
\end{equation}
both for massless fermions as well, and $D_l$ the lepton's diffusion constant. The different signs in Eq.(\ref{eq:coefficients}) with respect to the results obtained in Ref.~\cite{Cline:2020jre} come from having the expanding wall going to the left which means $\upsilon_{w}\to -\upsilon_{w}$.

The effective relaxation rate from inelastic scattering between
leptons and the bubble wall is,
\begin{equation}
\Gamma_{\mathrm{eff}}\equiv\left(\frac{1}{k_{L}}+\frac{1}{k_{R}}\right)\left(\Gamma_{m_{\tau}}+\Gamma_{Y_{\tau}}\right),
\end{equation}
with $\Gamma_{Y_{\tau}}=4.4\times10^{-7}\,T$~\cite{Cline:2021dkf}
and we calculate the chirality-flip rate from the $\tau$ thermal damping rate $\gamma_{\tau}\simeq 0.025\,T$~\cite{Elmfors:1998hh}, following the standard treatment of the flavor-source (VEV-resummed) framework~\cite{Lee:2004we,Cirigliano:2009yt,Chung:2009cb,DeVries:2018aul} (see also~\cite{deVries:2017ncy,Postma:2019scv,Huet:1995sh}), obtaining $\Gamma_{m_{\tau}}=0.3\,m_{\tau}^{2}/T\simeq 12\,(m_{\tau}^{2}/T^{2})\,\gamma_{\tau}$. We stress that $\Gamma_{m_\tau}$ is a CP-conserving relaxation rate, not a VIA CP-violating source term, and it is therefore unaffected by the demonstrated vanishing of the leading-order VIA source~\cite{Postma:2022dbr,vandeVis:2025qyz}.

The amplitude $\mathcal{A}$ in Eq.(\ref{eq:lepton_density}) is expressed as
\begin{equation}
\begin{aligned}\mathcal{A}\, =&  \intop_{0}^{\infty}dy\,\overline{S}_l(y)\,\frac{\,e^{-\gamma_{+}^{b}y}}{\overline{D}_l\,(\gamma_{+}^{b}-\gamma_{-}^{s})}\\
 & +\frac{1}{\Delta_s}\intop_{-2L_{w}}^{0}dy\,\overline{S}_l(y)\left[e^{-\gamma_{+}^{s}y}+\frac{\gamma_{-}^{s}-\gamma_{-}^{b}}{\gamma_{+}^{b}-\gamma_{-}^{s}}e^{-\gamma_{-}^{s}y}\right]\label{eq:Amplitude},
\end{aligned}
\end{equation}
where it is understood that we are approximating the collision rate
as a step function $\Gamma_{\mathrm{eff}}=\Gamma_{\mathrm{eff}}H(z)$ and $\gamma_{\pm}$ are the solutions of the auxiliary equation
of the homogeneous case for Eq.(\ref{eq:WKB_diffusion}), $\gamma_{\pm} =  1/2\overline{D}_l\,(v_{w}\pm\Delta_l)$, with the $b$ and $s$ notations referring to the
$\mathit{broken}$ and $\mathit{symmetric}$ phases respectively and $\Delta_l=\sqrt{v_{w}^{2}+4\,\overline{\Gamma}_l\,\overline{D}_l}$.






\section{Yukawa Lagrangian}
The most general Yukawa Lagrangian for leptons in the generic scalar basis $\{\Phi_{1},\Phi_{2}\}$ for the G2HDM  reads~\cite{Herrero-Garcia:2019mcy}:
\begin{equation}
    -\mathcal{L}_{Y}=\bar{L}_{L}^{0}\,(\tilde{Y}^{1}_{l}\Phi_{1}+\tilde{Y}^{2}_{l}\Phi_{2})l_{R}^{0}+{\rm ~h.c.}\,\label{eq:yuk2d},
\end{equation}
with $L=(\nu_l,\,l)^{T}$ and the ``0'' notation refers to the flavor eigenstates. The charged lepton masses are
\begin{equation}
    M_{l}=\frac{1}{\sqrt{2}}(v_{1}\tilde{Y}^{1}_{l}+v_{2}\tilde{Y}^{2}_{l}),\label{eq:mass-fermions}
\end{equation}
with $v_1$ and $v_2$ the VEVs of $\Phi_{1}$ and $\Phi_{2}$ respectively at zero temperature and $v=\sqrt{v_1^2+v_2^2}=246.22\,\mathrm{GeV}$.  Eq.(\ref{eq:mass-fermions}) is diagonalized through a bi-unitary transformation $\bar{M}_{l}=V_{lL}^{\dagger}M_{l}V_{lR}$ where the fact that $M_{l}$ is Hermitian implies that
$V_l \equiv V_{lL}=V_{lR}$, and the mass eigenstates for the charged leptons are given by $l=V_{l}^{\dagger}l^{0}$. In this way, Eq.(\ref{eq:mass-fermions}) takes the form $\bar{M}_{l}=1/\sqrt{2}(v_{1}Y^{1}_{l}+v_{2}Y^{2}_{l})$ where $Y^{i}_{l}=V_{lL}^{\dagger}\tilde{Y}^{i}_{l}V_{lR}$, diagonalizing then the charged lepton mass matrix. Using these expressions, we can write the Yukawa Lagrangian in the mass basis as,
\begin{equation}
-\mathcal{L}_{Y} =  \bar{\nu}_{b}\rho^{ba}_{\ell}P_{R}l_{a}\,H^{+}+\sum_{\phi}\bar{l}_{b} \Gamma^{\phi ba}_{l}P_{R}l_{a}\phi+\mathrm{h.c.},
\label{eq:YukawaLagrangian}
\end{equation}
for $\phi=h,H,A$, here $a,\,b$ are explicit flavor indices and

\begin{align}
\rho^{ba}_{l}& \equiv\dfrac{Y^{2,ba}_{l}}{\cos\beta}-\dfrac{\sqrt{2}\tan\beta\bar{M}_{l}^{ba}}{v},\nonumber\\
\Gamma^{hba}_{l} & \equiv\dfrac{\bar{M}_{l}^{ba}}{v}s_{\beta-\alpha}+\dfrac{1}{\sqrt{2}}\rho^{ba}_{l}c_{\beta-\alpha},\nonumber\\
\Gamma^{Hba}_{l} & \equiv\dfrac{\bar{M}_{l}^{ba}}{v}c_{\beta-\alpha}-\dfrac{1}{\sqrt{2}}\rho^{ba}_{l}s_{\beta-\alpha},\nonumber\\
\Gamma^{Aba}_{l} & \equiv\dfrac{i}{\sqrt{2}}\rho^{ba}_{l},\label{eq:Gammafphiba}
\end{align}
with $\tan\beta\equiv v_2/v_1$ at zero temperature and $\beta-\alpha$ is the mixing angle between the CP-even Higgs mass eigenstates $h,\,H$ relative to the Higgs basis.

\section{Source term in the G2HDM}
In the Higgs basis, we use a kink profile for
the first Higgs and an hyperbolic secant for the small non-zero second Higgs profile in the bubble wall~\cite{Kanemura:2023juv} defined by,
\begin{align}
h_{1}(z) & =\frac{v_{1}(T_n)}{2}\left[1+\tanh\left(\frac{z}{L_{w}}\right)\right],\nonumber\\
h_{2}(z) & =v_{2}(T_n)\,\textrm{sech}^2\left(\frac{z}{L_{w}}\right)\label{eq:kink},
\end{align}
where $v_{1,2}$ are the VEVs at the nucleation temperature $T_n$ and we have assumed both profiles have the same width $L_w$. The tunneling profiles at $T_n$ can also be used, introducing deviations relative to Eqs.(\ref{eq:kink}) though minor for most cases~\cite{Goncalves:2023svb}, especially when not considering CP-violation from the scalar sector. We consider the source of CP violation necessary for generating the BAU coming from the Yukawa lepton sector instead and promote the lepton mass matrix in Eq.(\ref{eq:mass-fermions}) to be $z$-dependent. We use a texture similar to the one implemented in the quark sector in \cite{Kanemura:2023juv} in the Higgs basis and set,
\begin{equation}
M_{l}(z)=\frac{1}{\sqrt{2}}\left[\left(\begin{array}{cc}
y_{\mu} & 0\\
0 & y_{\tau}
\end{array}\right)h_{1}+\left(\begin{array}{cc}
\rho_{\mu\mu} & \rho_{\mu\tau}\\
 \rho_{\tau\mu} & |\rho_{\tau\tau}|\,e^{i\theta}
\end{array}\right)h_{2}\right],
\label{eq:lepton_mass_texture}\end{equation}

 where $\rho_{ij}=\rho^{ij}_{\ell}$. The texture in Eq.(\ref{eq:lepton_mass_texture}) is also used in \cite{Crivellin:2023sig,Athron:2024rir} with an extra $\rho_{e\tau}$ coupling which does not affect significantly the $R(D^{(*)})$ ratios.  Thus our Yukawa textures here are consistent with the data as demonstrated  by the global fit results of Ref.~\cite{Athron:2024rir}, where in particular we consider the latest data in lepton flavor violating searches of $\tau\to\mu\gamma$, $\mu\to\,e\gamma$, $\tau\to\,3\mu$ and $h\to\,l\tau$ for $l=e,\,\mu$.  

Identically to Refs.~\cite{Kainulainen:2002th,Cline:2021dkf} in the case of charginos, in order to calculate the matrix
$A=UM_{l}\partial_{z}M_{l}^{-1}U^{\dagger}$, we diagonalize the $M_{l}^{\dagger}M_{l}$ matrix by the unitary transformation $U$~\cite{Cline:2001rk},
\begin{equation}
U=\frac{\sqrt{2}}{\sqrt{\Lambda(\Lambda+\Delta)}}\left(\begin{array}{cc}
\frac{1}{2}(\Lambda+\Delta) & a\\
-a^{*} & \frac{1}{2}(\Lambda+\Delta)
\end{array}\right),\label{eq:Utransformation}
\end{equation}
for $\Delta=(M_{l}^{\dagger}M_{l})_{11}-(M_{l}^{\dagger}M_{l})_{22}$,
$a=(M_{l}^{\dagger}M_{l})_{12}$ and $\Lambda=\sqrt{\Delta^{2}+4\left|a\right|^{2}}$. Expanding in small $\rho_{\mu\mu}$ and $\rho_{\mu\tau}=\rho_{\tau\mu}$ Yukawa couplings, the $A_{22}$ matrix element corresponding to the heaviest eigenstate is then,
\begin{equation}
A_{22}=\frac{y_{\tau}\,\rho_{\tau\tau}}{2}[h_{1}(z)h_{2}'(z)-h_{2}(z)h_{1}'(z)].\label{eq:A22}
\end{equation}

In this way, we find for the source term,

\begin{equation}
S_{j}=-\upsilon_{w}\gamma_{w}\,Q_{j}^{8o}\,({\rm Im}(A_{22}))'\,,\quad j=1,2,
\label{eq:tau_source}
\end{equation}

where $\gamma_w$ is the relativistic Lorentz factor and the functions $Q_{1,2}^{8o}$ are given in the massless case by
\begin{align}
Q_{2}^{8o}= & \frac{3}{2\pi^{2}\,\upsilon_{w}T^{2}\,\gamma_{w}^{2}}\tanh^{-1}\upsilon_{w},\nonumber\\
Q_{1}^{8o}= & -\upsilon_w\,Q_{2}^{8o}.
\end{align}
As in Ref.~\cite{Cline:2021dkf}, we have also neglected the contribution proportional to $Q_{j}^{9o}$ in Eq.(\ref{eq:tau_source}) which we have corroborated to be subdominant. We also note that Eq.(\ref{eq:tau_source}) can be written in terms of ${\rm Im}(J_A)=y_{\tau}\,|\rho_{\tau\tau}|\,\sin\theta$ which is the imaginary part of the Jarlskog invariant \cite{Jarlskog:1985ht,Botella:1994cs,Guo:2016ixx},
\begin{equation}
J_{A}=\frac{1}{\upsilon^{2}\mu_{12}^{\mathrm{HB}}}\sum_{a,b,c=1}^{2}\upsilon_{a}\upsilon_{b}^{*}\mu_{bc}\mathrel{\mathrm{Tr}[Y_{c}Y_{a}^{\dagger}],}\label{eq:Jarlskog_invariant}
\end{equation}

where $\mu_{ab}$ is the coefficient of the $\Phi_a\Phi_b$ term in the Higgs potential as defined in \cite{Guo:2016ixx}, $\mu_{12}^{\mathrm{HB}}$ is defined in the Higgs basis and the matrices $Y_{c,a}$ are given by the Yukawa textures in \ref{eq:lepton_mass_texture}.

\bibliographystyle{JHEP} 
\bibliography{GTHDM}

@article{Cline:2020jre,
    author = "Cline, James M. and Kainulainen, Kimmo",
    title = "{Electroweak baryogenesis at high bubble wall velocities}",
    eprint = "2001.00568",
    archivePrefix = "arXiv",
    primaryClass = "hep-ph",
    reportNumber = "CERN-TH-2019-227",
    doi = "10.1103/PhysRevD.101.063525",
    journal = "Phys. Rev. D",
    volume = "101",
    number = "6",
    pages = "063525",
    year = "2020"
}

@article{Cline:2021dkf,
    author = "Cline, James M. and Laurent, Benoit",
    title = "{Electroweak baryogenesis from light fermion sources: A critical study}",
    eprint = "2108.04249",
    archivePrefix = "arXiv",
    primaryClass = "hep-ph",
    doi = "10.1103/PhysRevD.104.083507",
    journal = "Phys. Rev. D",
    volume = "104",
    number = "8",
    pages = "083507",
    year = "2021"
}

@article{Kanemura:2023juv,
    author = "Kanemura, Shinya and Mura, Yushi",
    title = "{Electroweak baryogenesis via top-charm mixing}",
    eprint = "2303.11252",
    archivePrefix = "arXiv",
    primaryClass = "hep-ph",
    reportNumber = "OU-HET-1174",
    doi = "10.1007/JHEP09(2023)153",
    journal = "JHEP",
    volume = "09",
    pages = "153",
    year = "2023"
}

@article{Chupp:2017rkp,
    author = "Chupp, Timothy and Fierlinger, Peter and Ramsey-Musolf, Michael and Singh, Jaideep",
    title = "{Electric dipole moments of atoms, molecules, nuclei, and particles}",
    eprint = "1710.02504",
    archivePrefix = "arXiv",
    primaryClass = "physics.atom-ph",
    reportNumber = "ACFI-T17-18",
    doi = "10.1103/RevModPhys.91.015001",
    journal = "Rev. Mod. Phys.",
    volume = "91",
    number = "1",
    pages = "015001",
    year = "2019"
}

@article{Manton:1983nd,
    author = "Manton, N. S.",
    title = "{Topology in the Weinberg-Salam Theory}",
    reportNumber = "NSF-ITP-83-64",
    doi = "10.1103/PhysRevD.28.2019",
    journal = "Phys. Rev. D",
    volume = "28",
    pages = "2019",
    year = "1983"
}

@article{Klinkhamer:1984di,
    author = "Klinkhamer, Frans R. and Manton, N. S.",
    title = "{A Saddle Point Solution in the Weinberg-Salam Theory}",
    reportNumber = "NSF-ITP-84-57",
    doi = "10.1103/PhysRevD.30.2212",
    journal = "Phys. Rev. D",
    volume = "30",
    pages = "2212",
    year = "1984"
}

@article{Kuzmin:1985mm,
    author = "Kuzmin, V. A. and Rubakov, V. A. and Shaposhnikov, M. E.",
    title = "{On the Anomalous Electroweak Baryon Number Nonconservation in the Early Universe}",
    reportNumber = "IC/85/8",
    doi = "10.1016/0370-2693(85)91028-7",
    journal = "Phys. Lett. B",
    volume = "155",
    pages = "36",
    year = "1985"
}

@article{Caprini:2015zlo,
    author = "Caprini, Chiara and others",
    title = "{Science with the space-based interferometer eLISA. II: Gravitational waves from cosmological phase transitions}",
    eprint = "1512.06239",
    archivePrefix = "arXiv",
    primaryClass = "astro-ph.CO",
    reportNumber = "DESY-15-246",
    doi = "10.1088/1475-7516/2016/04/001",
    journal = "JCAP",
    volume = "04",
    pages = "001",
    year = "2016"
}

@article{Huang:2016odd,
    author = "Huang, Fa Peng and Wan, Youping and Wang, Dong-Gang and Cai, Yi-Fu and Zhang, Xinmin",
    title = "{Hearing the echoes of electroweak baryogenesis with gravitational wave detectors}",
    eprint = "1601.01640",
    archivePrefix = "arXiv",
    primaryClass = "hep-ph",
    doi = "10.1103/PhysRevD.94.041702",
    journal = "Phys. Rev. D",
    volume = "94",
    number = "4",
    pages = "041702",
    year = "2016"
}

@article{Ramsey-Musolf:2019lsf,
    author = "Ramsey-Musolf, Michael J.",
    title = "{The electroweak phase transition: a collider target}",
    eprint = "1912.07189",
    archivePrefix = "arXiv",
    primaryClass = "hep-ph",
    reportNumber = "ACFI-T19-14",
    doi = "10.1007/JHEP09(2020)179",
    journal = "JHEP",
    volume = "09",
    pages = "179",
    year = "2020"
}

@article{Shaposhnikov:1987tw,
    author = "Shaposhnikov, M. E.",
    title = "{Baryon Asymmetry of the Universe in Standard Electroweak Theory}",
    doi = "10.1016/0550-3213(87)90127-1",
    journal = "Nucl. Phys. B",
    volume = "287",
    pages = "757--775",
    year = "1987"
}

@article{Shaposhnikov:1986jp,
    author = "Shaposhnikov, M. E.",
    title = "{Possible Appearance of the Baryon Asymmetry of the Universe in an Electroweak Theory}",
    journal = "JETP Lett.",
    volume = "44",
    pages = "465--468",
    year = "1986"
}

@article{Curtin:2014jma,
    author = "Curtin, David and Meade, Patrick and Yu, Chiu-Tien",
    title = "{Testing Electroweak Baryogenesis with Future Colliders}",
    eprint = "1409.0005",
    archivePrefix = "arXiv",
    primaryClass = "hep-ph",
    reportNumber = "YITP-SB-14-33",
    doi = "10.1007/JHEP11(2014)127",
    journal = "JHEP",
    volume = "11",
    pages = "127",
    year = "2014"
}

@article{Beniwal:2017eik,
    author = "Beniwal, Ankit and Lewicki, Marek and Wells, James D. and White, Martin and Williams, Anthony G.",
    title = "{Gravitational wave, collider and dark matter signals from a scalar singlet electroweak baryogenesis}",
    eprint = "1702.06124",
    archivePrefix = "arXiv",
    primaryClass = "hep-ph",
    reportNumber = "ADP-17-08-T1014, ADP--17--08-T1014",
    doi = "10.1007/JHEP08(2017)108",
    journal = "JHEP",
    volume = "08",
    pages = "108",
    year = "2017"
}

@inproceedings{Cline:2006ts,
    author = "Cline, James M.",
    title = "{Baryogenesis}",
    booktitle = "{Les Houches Summer School - Session 86: Particle Physics and Cosmology: The Fabric of Spacetime}",
    eprint = "hep-ph/0609145",
    archivePrefix = "arXiv",
    month = "9",
    year = "2006"
}

@article{Cline:2001rk,
    author = "Cline, James M. and Joyce, Michael and Kainulainen, Kimmo",
    title = "{Supersymmetric electroweak baryogenesis}",
    eprint = "hep-ph/0006119",
    archivePrefix = "arXiv",
    reportNumber = "MCGILL-00-15, NORDITA-2000-38-HE, LPT-ORSAY-00-46",
    doi = "10.1088/1126-6708/2000/07/018",
    journal = "JHEP",
    volume = "07",
    pages = "018",
    year = "2000"
}

@article{DeVries:2018aul,
    author = "De Vries, Jordy and Postma, Marieke and van de Vis, Jorinde",
    title = "{The role of leptons in electroweak baryogenesis}",
    eprint = "1811.11104",
    archivePrefix = "arXiv",
    primaryClass = "hep-ph",
    reportNumber = "Nikhef-2018-056",
    doi = "10.1007/JHEP04(2019)024",
    journal = "JHEP",
    volume = "04",
    pages = "024",
    year = "2019"
}

@article{Chung:2009cb,
    author = "Chung, Daniel J. H. and Garbrecht, Bjorn and Ramsey-Musolf, Michael J. and Tulin, Sean",
    title = "{Lepton-mediated electroweak baryogenesis}",
    eprint = "0905.4509",
    archivePrefix = "arXiv",
    primaryClass = "hep-ph",
    reportNumber = "NPAC-09-05",
    doi = "10.1103/PhysRevD.81.063506",
    journal = "Phys. Rev. D",
    volume = "81",
    pages = "063506",
    year = "2010"
}

@article{Lee:2004we,
    author = "Lee, Christopher and Cirigliano, Vincenzo and Ramsey-Musolf, Michael J.",
    title = "{Resonant relaxation in electroweak baryogenesis}",
    eprint = "hep-ph/0412354",
    archivePrefix = "arXiv",
    reportNumber = "CALTECH-MAP-304, CALT-68-2535",
    doi = "10.1103/PhysRevD.71.075010",
    journal = "Phys. Rev. D",
    volume = "71",
    pages = "075010",
    year = "2005"
}

@article{Xie:2020wzn,
    author = "Xie, Ke-Pan",
    title = "{Lepton-mediated electroweak baryogenesis, gravitational waves and the $4\tau$ final state at the collider}",
    eprint = "2011.04821",
    archivePrefix = "arXiv",
    primaryClass = "hep-ph",
    doi = "10.1007/JHEP02(2021)090",
    journal = "JHEP",
    volume = "02",
    pages = "090",
    year = "2021",
    note = "[Erratum: JHEP 8, 052 (2022)]"
}

@article{Cirigliano:2006wh,
    author = "Cirigliano, Vincenzo and Ramsey-Musolf, Michael J. and Tulin, Sean and Lee, Christopher",
    title = "{Yukawa and tri-scalar processes in electroweak baryogenesis}",
    eprint = "hep-ph/0603058",
    archivePrefix = "arXiv",
    reportNumber = "CALTECH-MAP-312, INT-PUB-06-03",
    doi = "10.1103/PhysRevD.73.115009",
    journal = "Phys. Rev. D",
    volume = "73",
    pages = "115009",
    year = "2006"
}

@article{Li:2024mts,
    author = "Li, Yuan-Zhen and Ramsey-Musolf, Michael J. and Yu, Jiang-Hao",
    title = "{Does the Electron EDM Preclude Electroweak Baryogenesis ?}",
    eprint = "2404.19197",
    archivePrefix = "arXiv",
    primaryClass = "hep-ph",
    month = "4",
    year = "2024"
}

@article{Cirigliano:2009yt,
    author = "Cirigliano, Vincenzo and Lee, Christopher and Ramsey-Musolf, Michael J. and Tulin, Sean",
    title = "{Flavored Quantum Boltzmann Equations}",
    eprint = "0912.3523",
    archivePrefix = "arXiv",
    primaryClass = "hep-ph",
    reportNumber = "NPAC-09-16, UCB-PTH-09-37",
    doi = "10.1103/PhysRevD.81.103503",
    journal = "Phys. Rev. D",
    volume = "81",
    pages = "103503",
    year = "2010"
}

@article{Enomoto:2022rrl,
    author = "Enomoto, Kazuki and Kanemura, Shinya and Mura, Yushi",
    title = "{New benchmark scenarios of electroweak baryogenesis in aligned two Higgs double models}",
    eprint = "2207.00060",
    archivePrefix = "arXiv",
    primaryClass = "hep-ph",
    reportNumber = "OU-HET-1147",
    doi = "10.1007/JHEP09(2022)121",
    journal = "JHEP",
    volume = "09",
    pages = "121",
    year = "2022"
}

@article{Sakharov:1967dj,
    author = "Sakharov, A. D.",
    title = "{Violation of CP Invariance, C asymmetry, and baryon asymmetry of the universe}",
    doi = "10.1070/PU1991v034n05ABEH002497",
    journal = "Pisma Zh. Eksp. Teor. Fiz.",
    volume = "5",
    pages = "32--35",
    year = "1967"
}

@article{Elmfors:1998hh,
    author = "Elmfors, Per and Enqvist, Kari and Riotto, Antonio and Vilja, Iiro",
    title = "{Damping rates in the MSSM and electroweak baryogenesis}",
    eprint = "hep-ph/9809529",
    archivePrefix = "arXiv",
    reportNumber = "HIP-1998-64-TH, SU-ITP-98-18, TURKU-FL-P30-98, CERN-TH-98-313",
    doi = "10.1016/S0370-2693(99)00169-0",
    journal = "Phys. Lett. B",
    volume = "452",
    pages = "279--286",
    year = "1999"
}

@article{Chiang:2016vgf,
    author = "Chiang, Cheng-Wei and Fuyuto, Kaori and Senaha, Eibun",
    title = "{Electroweak Baryogenesis with Lepton Flavor Violation}",
    eprint = "1607.07316",
    archivePrefix = "arXiv",
    primaryClass = "hep-ph",
    doi = "10.1016/j.physletb.2016.09.052",
    journal = "Phys. Lett. B",
    volume = "762",
    pages = "315--320",
    year = "2016"
}

@article{Alonso-Gonzalez:2021jsa,
    author = "Alonso-Gonz\'alez, J. and Merlo, L. and Pokorski, S.",
    title = "{A new bound on CP violation in the \ensuremath{\tau} lepton Yukawa coupling and electroweak baryogenesis}",
    eprint = "2103.16569",
    archivePrefix = "arXiv",
    primaryClass = "hep-ph",
    doi = "10.1007/JHEP06(2021)166",
    journal = "JHEP",
    volume = "06",
    pages = "166",
    year = "2021"
}

@article{Kainulainen:2002th,
    author = "Kainulainen, Kimmo and Prokopec, Tomislav and Schmidt, Michael G. and Weinstock, Steffen",
    title = "{Semiclassical force for electroweak baryogenesis: Three-dimensional derivation}",
    eprint = "hep-ph/0202177",
    archivePrefix = "arXiv",
    reportNumber = "CERN-TH-2002-011, NORDITA-2002-5-HE, HD-THEP-02-7",
    doi = "10.1103/PhysRevD.66.043502",
    journal = "Phys. Rev. D",
    volume = "66",
    pages = "043502",
    year = "2002"
}

@article{Huet:1995sh,
    author = "Huet, Patrick and Nelson, Ann E.",
    title = "{Electroweak baryogenesis in supersymmetric models}",
    eprint = "hep-ph/9506477",
    archivePrefix = "arXiv",
    reportNumber = "UW-PT-95-07",
    doi = "10.1103/PhysRevD.53.4578",
    journal = "Phys. Rev. D",
    volume = "53",
    pages = "4578--4597",
    year = "1996"
}

@article{Cirigliano:2009yd,
    author = "Cirigliano, Vincenzo and Li, Yingchuan and Profumo, Stefano and Ramsey-Musolf, Michael J.",
    title = "{MSSM Baryogenesis and Electric Dipole Moments: An Update on the Phenomenology}",
    eprint = "0910.4589",
    archivePrefix = "arXiv",
    primaryClass = "hep-ph",
    reportNumber = "NPAC-09-12",
    doi = "10.1007/JHEP01(2010)002",
    journal = "JHEP",
    volume = "01",
    pages = "002",
    year = "2010"
}

@article{Jarlskog:1985ht,
    author = "Jarlskog, C.",
    title = "{Commutator of the Quark Mass Matrices in the Standard Electroweak Model and a Measure of Maximal CP Nonconservation}",
    reportNumber = "USIP-85-14",
    doi = "10.1103/PhysRevLett.55.1039",
    journal = "Phys. Rev. Lett.",
    volume = "55",
    pages = "1039",
    year = "1985"
}

@article{Botella:1994cs,
    author = "Botella, F. J. and Silva, Joao P.",
    title = "{Jarlskog - like invariants for theories with scalars and fermions}",
    eprint = "hep-ph/9411288",
    archivePrefix = "arXiv",
    reportNumber = "FTUV-94-68, IFIC-94-65",
    doi = "10.1103/PhysRevD.51.3870",
    journal = "Phys. Rev. D",
    volume = "51",
    pages = "3870--3875",
    year = "1995"
}

@article{Cirigliano:2011di,
    author = "Cirigliano, Vincenzo and Lee, Christopher and Tulin, Sean",
    title = "{Resonant Flavor Oscillations in Electroweak Baryogenesis}",
    eprint = "1106.0747",
    archivePrefix = "arXiv",
    primaryClass = "hep-ph",
    reportNumber = "MIT-CTP-4269",
    doi = "10.1103/PhysRevD.84.056006",
    journal = "Phys. Rev. D",
    volume = "84",
    pages = "056006",
    year = "2011"
}

@article{Cline:2011mm,
    author = "Cline, James M. and Kainulainen, Kimmo and Trott, Michael",
    title = "{Electroweak Baryogenesis in Two Higgs Doublet Models and B meson anomalies}",
    eprint = "1107.3559",
    archivePrefix = "arXiv",
    primaryClass = "hep-ph",
    doi = "10.1007/JHEP11(2011)089",
    journal = "JHEP",
    volume = "11",
    pages = "089",
    year = "2011"
}

@article{Goncalves:2023svb,
    author = "Gon\c{c}alves, Dorival and Kaladharan, Ajay and Wu, Yongcheng",
    title = "{Gravitational waves, bubble profile, and baryon asymmetry in the complex 2HDM}",
    eprint = "2307.03224",
    archivePrefix = "arXiv",
    primaryClass = "hep-ph",
    doi = "10.1103/PhysRevD.108.075010",
    journal = "Phys. Rev. D",
    volume = "108",
    number = "7",
    pages = "075010",
    year = "2023"
}

@article{Enomoto:2021dkl,
    author = "Enomoto, Kazuki and Kanemura, Shinya and Mura, Yushi",
    title = "{Electroweak baryogenesis in aligned two Higgs doublet models}",
    eprint = "2111.13079",
    archivePrefix = "arXiv",
    primaryClass = "hep-ph",
    reportNumber = "OU-HET-1114",
    doi = "10.1007/JHEP01(2022)104",
    journal = "JHEP",
    volume = "01",
    pages = "104",
    year = "2022"
}

@article{Dorsch:2016nrg,
    author = "Dorsch, G. C. and Huber, S. J. and Konstandin, T. and No, J. M.",
    title = "{A Second Higgs Doublet in the Early Universe: Baryogenesis and Gravitational Waves}",
    eprint = "1611.05874",
    archivePrefix = "arXiv",
    primaryClass = "hep-ph",
    reportNumber = "DESY-16-213",
    doi = "10.1088/1475-7516/2017/05/052",
    journal = "JCAP",
    volume = "05",
    pages = "052",
    year = "2017"
}

@article{Fromme:2006wx,
    author = "Fromme, Lars and Huber, Stephan J.",
    title = "{Top transport in electroweak baryogenesis}",
    eprint = "hep-ph/0604159",
    archivePrefix = "arXiv",
    reportNumber = "CERN-PH-TH-2006-064, BI-TP-2006-10",
    doi = "10.1088/1126-6708/2007/03/049",
    journal = "JHEP",
    volume = "03",
    pages = "049",
    year = "2007"
}

@article{Kainulainen:2021oqs,
    author = "Kainulainen, Kimmo",
    title = "{CP-violating transport theory for electroweak baryogenesis with thermal corrections}",
    eprint = "2108.08336",
    archivePrefix = "arXiv",
    primaryClass = "hep-ph",
    doi = "10.1088/1475-7516/2021/11/042",
    journal = "JCAP",
    volume = "11",
    number = "11",
    pages = "042",
    year = "2021"
}

@article{Fromme:2006cm,
    author = "Fromme, Lars and Huber, Stephan J. and Seniuch, Michael",
    title = "{Baryogenesis in the two-Higgs doublet model}",
    eprint = "hep-ph/0605242",
    archivePrefix = "arXiv",
    reportNumber = "CERN-PH-TH-2006-094, BI-TP-2006-18",
    doi = "10.1088/1126-6708/2006/11/038",
    journal = "JHEP",
    volume = "11",
    pages = "038",
    year = "2006"
}

@article{Basler:2021kgq,
    author = {Basler, Philipp and Biermann, Lisa and M\"uhlleitner, Margarete and M\"uller, Jonas},
    title = "{Electroweak baryogenesis in the CP-violating two-Higgs doublet model}",
    eprint = "2108.03580",
    archivePrefix = "arXiv",
    primaryClass = "hep-ph",
    reportNumber = "KA-TP-14-2021",
    doi = "10.1140/epjc/s10052-023-11192-9",
    journal = "Eur. Phys. J. C",
    volume = "83",
    number = "1",
    pages = "57",
    year = "2023"
}

@article{deVries:2017ncy,
    author = "de Vries, Jordy and Postma, Marieke and van de Vis, Jorinde and White, Graham",
    title = "{Electroweak Baryogenesis and the Standard Model Effective Field Theory}",
    eprint = "1710.04061",
    archivePrefix = "arXiv",
    primaryClass = "hep-ph",
    reportNumber = "Nikhef-2017-044",
    doi = "10.1007/JHEP01(2018)089",
    journal = "JHEP",
    volume = "01",
    pages = "089",
    year = "2018"
}

@article{Postma:2019scv,
    author = "Postma, Marieke and van de Vis, Jorinde",
    title = "{Source terms for electroweak baryogenesis in the vev-insertion approximation beyond leading order}",
    eprint = "1910.11794",
    archivePrefix = "arXiv",
    primaryClass = "hep-ph",
    reportNumber = "DESY-19-182",
    doi = "10.1007/JHEP02(2020)090",
    journal = "JHEP",
    volume = "02",
    pages = "090",
    year = "2020"
}

@article{Postma:2022dbr,
    author = "Postma, Marieke and van de Vis, Jorinde and White, Graham",
    title = "{Resummation and cancellation of the VIA source in electroweak baryogenesis}",
    eprint = "2206.01120",
    archivePrefix = "arXiv",
    primaryClass = "hep-ph",
    reportNumber = "Nikhef 2022-007, DESY-22-092, IPMU22-0034",
    doi = "10.1007/JHEP12(2022)121",
    journal = "JHEP",
    volume = "12",
    pages = "121",
    year = "2022"
}

@article{Guo:2016ixx,
    author = "Guo, Huai-Ke and Li, Ying-Ying and Liu, Tao and Ramsey-Musolf, Michael and Shu, Jing",
    title = "{Lepton-Flavored Electroweak Baryogenesis}",
    eprint = "1609.09849",
    archivePrefix = "arXiv",
    primaryClass = "hep-ph",
    doi = "10.1103/PhysRevD.96.115034",
    journal = "Phys. Rev. D",
    volume = "96",
    number = "11",
    pages = "115034",
    year = "2017"
}

@article{Arnold:2000dr,
    author = "Arnold, Peter Brockway and Moore, Guy D. and Yaffe, Laurence G.",
    title = "{Transport coefficients in high temperature gauge theories. 1. Leading log results}",
    eprint = "hep-ph/0010177",
    archivePrefix = "arXiv",
    reportNumber = "UW-PT-00-15",
    doi = "10.1088/1126-6708/2000/11/001",
    journal = "JHEP",
    volume = "11",
    pages = "001",
    year = "2000"
}

@article{Tulin:2011wi,
    author = "Tulin, Sean and Winslow, Peter",
    title = "{Anomalous $B$ meson mixing and baryogenesis}",
    eprint = "1105.2848",
    archivePrefix = "arXiv",
    primaryClass = "hep-ph",
    doi = "10.1103/PhysRevD.84.034013",
    journal = "Phys. Rev. D",
    volume = "84",
    pages = "034013",
    year = "2011"
}

@article{Fuyuto:2017ewj,
    author = "Fuyuto, Kaori and Hou, Wei-Shu and Senaha, Eibun",
    title = "{Electroweak baryogenesis driven by extra top Yukawa couplings}",
    eprint = "1705.05034",
    archivePrefix = "arXiv",
    primaryClass = "hep-ph",
    reportNumber = "ACFI-T17-12",
    doi = "10.1016/j.physletb.2017.11.073",
    journal = "Phys. Lett. B",
    volume = "776",
    pages = "402--406",
    year = "2018"
}

@article{Liu:2011jh,
    author = "Liu, Tao and Ramsey-Musolf, Michael J. and Shu, Jing",
    title = "{Electroweak Beautygenesis: From b$\to$s CP-violation to the Cosmic Baryon Asymmetry}",
    eprint = "1109.4145",
    archivePrefix = "arXiv",
    primaryClass = "hep-ph",
    reportNumber = "IPMU11-0076, NPAC-11-10, SISSA-48-2011-EP",
    doi = "10.1103/PhysRevLett.108.221301",
    journal = "Phys. Rev. Lett.",
    volume = "108",
    pages = "221301",
    year = "2012"
}

@article{Planck:2013pxb,
    author = "Ade, P. A. R. and others",
    collaboration = "Planck",
    title = "{Planck 2013 results. XVI. Cosmological parameters}",
    eprint = "1303.5076",
    archivePrefix = "arXiv",
    primaryClass = "astro-ph.CO",
    reportNumber = "CERN-PH-TH-2013-129",
    doi = "10.1051/0004-6361/201321591",
    journal = "Astron. Astrophys.",
    volume = "571",
    pages = "A16",
    year = "2014"
}

@article{Crivellin:2023sig,
    author = "Crivellin, Andreas and Iguro, Syuhei",
    title = "{Accumulating hints for flavor-violating Higgs bosons at the electroweak scale}",
    eprint = "2311.03430",
    archivePrefix = "arXiv",
    primaryClass = "hep-ph",
    reportNumber = "PSI-PR-23-41, ZU-TH 73/23, P3H-23-078, TTP23-051",
    doi = "10.1103/PhysRevD.110.015014",
    journal = "Phys. Rev. D",
    volume = "110",
    number = "1",
    pages = "015014",
    year = "2024"
}

@article{Blanke:2018yud,
    author = "Blanke, Monika and Crivellin, Andreas and de Boer, Stefan and Kitahara, Teppei and Moscati, Marta and Nierste, Ulrich and Ni\v{s}and\v{z}i\'c, Ivan",
    title = "{Impact of polarization observables and $ B_c\to \tau \nu$ on new physics explanations of the $b\to c \tau \nu$ anomaly}",
    eprint = "1811.09603",
    archivePrefix = "arXiv",
    primaryClass = "hep-ph",
    reportNumber = "PSI-PR-18-16; TTP-18-42, PSI-PR--18--16, TTP--18--42",
    doi = "10.1103/PhysRevD.99.075006",
    journal = "Phys. Rev. D",
    volume = "99",
    number = "7",
    pages = "075006",
    year = "2019"
}

@article{Athron:2021auq,
    author = "Athron, Peter and Balazs, Csaba and Gonzalo, Tom\'as E. and Jacob, Douglas and Mahmoudi, Farvah and Sierra, Cristian",
    title = "{Likelihood analysis of the flavour anomalies and g \textendash{} 2 in the general two Higgs doublet model}",
    eprint = "2111.10464",
    archivePrefix = "arXiv",
    primaryClass = "hep-ph",
    reportNumber = "CERN-TH-2021-194, TTK-21-47",
    doi = "10.1007/JHEP01(2022)037",
    journal = "JHEP",
    volume = "01",
    pages = "037",
    year = "2022"
}

@article{Athron:2021jyv,
    author = "Athron, Peter and Balazs, Csaba and Gonzalo, Tom\'as E. and Jacob, Douglas and Mahmoudi, Farvah and Sierra, Cristian",
    title = "{Likelihood analysis of the flavour anomalies and $g-2$ in the general two Higgs doublet model}",
    eprint = "2111.10464",
    archivePrefix = "arXiv",
    primaryClass = "hep-ph",
    reportNumber = "CERN-TH-2021-194, TTK-21-47",
    month = "11",
    year = "2021"
}

@article{HFLAV:2022esi,
    author = "Amhis, Yasmine Sara and others",
    collaboration = "HFLAV",
    title = "{Averages of b-hadron, c-hadron, and \ensuremath{\tau}-lepton properties as of 2021}",
    eprint = "2206.07501",
    archivePrefix = "arXiv",
    primaryClass = "hep-ex",
    doi = "10.1103/PhysRevD.107.052008",
    journal = "Phys. Rev. D",
    volume = "107",
    number = "5",
    pages = "052008",
    year = "2023"
}

@article{ATLAS:2023bzb,
    author = "Aad, Georges and others",
    collaboration = "ATLAS",
    title = "{Search for a light charged Higgs boson in $t \rightarrow H^{\pm}b$ decays, with $H^{\pm} \rightarrow cb$, in the lepton+jets final state in proton-proton collisions at $\sqrt{s}=13$ TeV with the ATLAS detector}",
    eprint = "2302.11739",
    archivePrefix = "arXiv",
    primaryClass = "hep-ex",
    reportNumber = "CERN-EP-2022-207",
    doi = "10.1007/JHEP09(2023)004",
    journal = "JHEP",
    volume = "09",
    pages = "004",
    year = "2023"
}

@article{Herrero-Garcia:2019mcy,
    author = "Herrero-Garcia, Juan and Nebot, Miguel and Rajec, Filip and White, Martin and Williams, Anthony G.",
    archivePrefix = "arXiv",
    doi = "10.1007/JHEP02(2020)147",
    eprint = "1907.05900",
    journal = "JHEP",
    pages = "147",
    primaryClass = "hep-ph",
    title = "{Higgs Quark Flavor Violation: Simplified Models and Status of General Two-Higgs-Doublet Model}",
    volume = "02",
    year = "2020"
}

@article{Crivellin:2023zui,
    author = "Crivellin, Andreas and Mellado, Bruce",
    title = "{Anomalies in Particle Physics}",
    eprint = "2309.03870",
    archivePrefix = "arXiv",
    primaryClass = "hep-ph",
    reportNumber = "PSI-PR-23-34, ZU-TH 53/23, ICPP-73",
    doi = "10.1038/s42254-024-00703-6",
    journal = "Nature Rev.Phys.",
    year = "2024"
}

@article{Alguero:2023jeh,
    author = "Alguer\'o, Marcel and Biswas, Aritra and Capdevila, Bernat and Descotes-Genon, S\'ebastien and Matias, Joaquim and Novoa-Brunet, Mart\'\i{}n",
    title = "{To (b)e or not to (b)e: no electrons at LHCb}",
    eprint = "2304.07330",
    archivePrefix = "arXiv",
    primaryClass = "hep-ph",
    reportNumber = "BARI-TH/23-747",
    doi = "10.1140/epjc/s10052-023-11824-0",
    journal = "Eur. Phys. J. C",
    volume = "83",
    number = "7",
    pages = "648",
    year = "2023"
}

@article{Greljo:2022jac,
    author = "Greljo, Admir and Salko, Jakub and Smolkovi\v{c}, Aleks and Stangl, Peter",
    title = "{Rare b decays meet high-mass Drell-Yan}",
    eprint = "2212.10497",
    archivePrefix = "arXiv",
    primaryClass = "hep-ph",
    reportNumber = "CERN-TH-2023-037",
    doi = "10.1007/JHEP05(2023)087",
    journal = "JHEP",
    volume = "05",
    pages = "087",
    year = "2023"
}

@article{Hou:2023ran,
    author = "Hou, Wei-Shu and Kumar, Girish and Modak, Tanmoy",
    title = "{Probing baryogenesis with radiative beauty decay and electron electric dipole moment}",
    eprint = "2302.08847",
    archivePrefix = "arXiv",
    primaryClass = "hep-ph",
    doi = "10.1103/PhysRevD.109.L011701",
    journal = "Phys. Rev. D",
    volume = "109",
    number = "1",
    pages = "L011701",
    year = "2024"
}

@article{Nelson:2019fln,
    author = "Nelson, Ann E. and Xiao, Huangyu",
    title = "{Baryogenesis from B Meson Oscillations}",
    eprint = "1901.08141",
    archivePrefix = "arXiv",
    primaryClass = "hep-ph",
    doi = "10.1103/PhysRevD.100.075002",
    journal = "Phys. Rev. D",
    volume = "100",
    number = "7",
    pages = "075002",
    year = "2019"
}

@article{Chung:2008aya,
    author = "Chung, Daniel J. H. and Garbrecht, Bjorn and Ramsey-Musolf, Michael J. and Tulin, Sean",
    title = "{Yukawa Interactions and Supersymmetric Electroweak Baryogenesis}",
    eprint = "0808.1144",
    archivePrefix = "arXiv",
    primaryClass = "hep-ph",
    doi = "10.1103/PhysRevLett.102.061301",
    journal = "Phys. Rev. Lett.",
    volume = "102",
    pages = "061301",
    year = "2009"
}

@article{Martinez:2018ynq,
    author = "Martinez, R. and Sierra, C.F. and Valencia, German",
    title = "{Beyond $\mathcal{R}(D^{(*)})$ with the general type-III 2HDM for $b\to c\tau\nu$}",
    eprint = "1805.04098",
    archivePrefix = "arXiv",
    primaryClass = "hep-ph",
    doi = "10.1103/PhysRevD.98.115012",
    journal = "Phys. Rev. D",
    volume = "98",
    number = "11",
    pages = "115012",
    year = "2018"
}

@article{Bhom:2020bfe,
    author = "Bhom, Jihyun and Chrzaszcz, Marcin",
    title = "{HEPLike: an open source framework for experimental likelihood evaluation}",
    eprint = "2003.03956",
    archivePrefix = "arXiv",
    primaryClass = "physics.data-an",
    doi = "10.1016/j.cpc.2020.107235",
    journal = "Comput. Phys. Commun.",
    volume = "254",
    pages = "107235",
    year = "2020"
}

@article{ParticleDataGroup:2024cfk,
    author = "Navas, S. and others",
    collaboration = "Particle Data Group",
    title = "{Review of particle physics}",
    doi = "10.1103/PhysRevD.110.030001",
    journal = "Phys. Rev. D",
    volume = "110",
    number = "3",
    pages = "030001",
    year = "2024"
}

@techreport{CMS-PAS-BPH-21-002,
      collaboration = "CMS",
      title         = "{Angular analysis of the $B^0 \to K^{*0}(892) \mu^+ \mu^-$
                       decay at $\sqrt{s}$ = 13 TeV}",
      institution   = "CERN",
      reportNumber  = "CMS-PAS-BPH-21-002",
      address       = "Geneva",
      year          = "2024",
      url           = "https://cds.cern.ch/record/2899589",
}

@article{LHCb:2020lmf,
    author = "Aaij, Roel and others",
    collaboration = "LHCb",
    title = "{Measurement of $CP$-Averaged Observables in the $B^{0}\rightarrow K^{*0}\mu^{+}\mu^{-}$ Decay}",
    eprint = "2003.04831",
    archivePrefix = "arXiv",
    primaryClass = "hep-ex",
    reportNumber = "LHCb-PAPER-2020-002, CERN-EP-2020-027",
    doi = "10.1103/PhysRevLett.125.011802",
    journal = "Phys. Rev. Lett.",
    volume = "125",
    number = "1",
    pages = "011802",
    year = "2020"
}

@article{Mahmoudi:2007vz,
    author = "Mahmoudi, F.",
    title = "{SuperIso: A Program for calculating the isospin asymmetry of B $\to$K* gamma in the MSSM}",
    eprint = "0710.2067",
    archivePrefix = "arXiv",
    primaryClass = "hep-ph",
    doi = "10.1016/j.cpc.2007.12.006",
    journal = "Comput. Phys. Commun.",
    volume = "178",
    pages = "745--754",
    year = "2008"
}

@article{Mahmoudi:2008tp,
    author = "Mahmoudi, F.",
    title = "{SuperIso v2.3: A Program for calculating flavor physics observables in Supersymmetry}",
    eprint = "0808.3144",
    archivePrefix = "arXiv",
    primaryClass = "hep-ph",
    doi = "10.1016/j.cpc.2009.02.017",
    journal = "Comput. Phys. Commun.",
    volume = "180",
    pages = "1579--1613",
    year = "2009"
}

@article{Mahmoudi:2009zz,
    author = "Mahmoudi, F.",
    title = "{SuperIso v3.0, flavor physics observables calculations: Extension to NMSSM}",
    doi = "10.1016/j.cpc.2009.05.001",
    journal = "Comput. Phys. Commun.",
    volume = "180",
    pages = "1718--1719",
    year = "2009"
}

@article{Athron:2017ard,
    author = "Athron, Peter and others",
    collaboration = "GAMBIT Collaboration",
    title = "{GAMBIT: The Global and Modular Beyond-the-Standard-Model Inference Tool}",
    eprint = "1705.07908",
    archivePrefix = "arXiv",
    primaryClass = "hep-ph",
    reportNumber = "COEPP-MN-17-6, CERN-TH-2017-166, NORDITA-2017-074, DESY-17-236",
    doi = "10.1140/epjc/s10052-017-5321-8",
    journal = "Eur. Phys. J. C",
    volume = "77",
    number = "11",
    pages = "784",
    year = "2017",
    note = "[Addendum: Eur.Phys.J.C 78, 98 (2018)]."
}

@article{GAMBITModelsWorkgroup:2017ilg,
    author = "Athron, Peter and others",
    collaboration = "GAMBIT Models Workgroup",
    title = "{SpecBit, DecayBit and PrecisionBit: GAMBIT modules for computing mass spectra, particle decay rates and precision observables}",
    eprint = "1705.07936",
    archivePrefix = "arXiv",
    primaryClass = "hep-ph",
    reportNumber = "COEPP-MN-17-7, CoEPP-MN-17-7, gambit-code, gambit-code-2017",
    doi = "10.1140/epjc/s10052-017-5390-8",
    journal = "Eur. Phys. J. C",
    volume = "78",
    number = "1",
    pages = "22",
    year = "2018"
}

@article{Kamenik:2008tj,
    author = "Kamenik, Jernej F. and Mescia, Federico",
    title = "{$B \to  D \tau \nu$ Branching Ratios: Opportunity for Lattice QCD and Hadron Colliders}",
    eprint = "0802.3790",
    archivePrefix = "arXiv",
    primaryClass = "hep-ph",
    doi = "10.1103/PhysRevD.78.014003",
    journal = "Phys. Rev. D",
    volume = "78",
    pages = "014003",
    year = "2008"
}

@article{deDivitiis:2007ptj,
    author = "de Divitiis, G. M. and Petronzio, R. and Tantalo, N.",
    title = "{Quenched lattice calculation of semileptonic heavy-light meson form factors}",
    eprint = "0707.0587",
    archivePrefix = "arXiv",
    primaryClass = "hep-lat",
    doi = "10.1088/1126-6708/2007/10/062",
    journal = "JHEP",
    volume = "10",
    pages = "062",
    year = "2007"
}

@misc{HFLAV2024winter,
   collaboration = "HFLAV",
  howpublished = "{``Preliminary average of $R(D)$ and $R(D^\ast)$ for Moriond 2024'' at \url{https://hflav-eos.web.cern.ch/hflav-eos/semi/moriond24/html/RDsDsstar/RDRDs.html}}"
}

@article{Eriksson:2009ws,
    author = "Eriksson, David and Rathsman, Johan and Stal, Oscar",
    title = "{2HDMC: Two-Higgs-Doublet Model Calculator Physics and Manual}",
    eprint = "0902.0851",
    archivePrefix = "arXiv",
    primaryClass = "hep-ph",
    doi = "10.1016/j.cpc.2009.09.011",
    journal = "Comput. Phys. Commun.",
    volume = "181",
    pages = "189--205",
    year = "2010"
}

@article{Cline:1995dg,
    author = "Cline, James M. and Kainulainen, Kimmo and Vischer, Axel P.",
    title = "{Dynamics of two Higgs doublet CP violation and baryogenesis at the electroweak phase transition}",
    eprint = "hep-ph/9506284",
    archivePrefix = "arXiv",
    reportNumber = "MCGILL-95-16, CERN-TH-95-136, UMN-TH-1343-94",
    doi = "10.1103/PhysRevD.54.2451",
    journal = "Phys. Rev. D",
    volume = "54",
    pages = "2451--2472",
    year = "1996"
}

@article{Cohen:1991iu,
    author = "Cohen, Andrew G. and Kaplan, D. B. and Nelson, A. E.",
    title = "{Spontaneous baryogenesis at the weak phase transition}",
    reportNumber = "UCSD-PTH-91-11, BUHEP-91-5",
    doi = "10.1016/0370-2693(91)91711-4",
    journal = "Phys. Lett. B",
    volume = "263",
    pages = "86--92",
    year = "1991"
}

@article{Fuchs:2020uoc,
    author = "Fuchs, Elina and Losada, Marta and Nir, Yosef and Viernik, Yehonatan",
    title = "{$CP$ violation from $\tau$, $t$ and $b$ dimension-6 Yukawa couplings - interplay of baryogenesis, EDM and Higgs physics}",
    eprint = "2003.00099",
    archivePrefix = "arXiv",
    primaryClass = "hep-ph",
    reportNumber = "FERMILAB-PUB-20-072-T, EFI-20-3",
    doi = "10.1007/JHEP05(2020)056",
    journal = "JHEP",
    volume = "05",
    pages = "056",
    year = "2020"
}

@article{Ge:2020mcl,
    author = "Ge, Shao-Feng and Li, Gang and Pasquini, Pedro and Ramsey-Musolf, Michael J.",
    title = "{CP-violating Higgs Di-tau Decays: Baryogenesis and Higgs Factories}",
    eprint = "2012.13922",
    archivePrefix = "arXiv",
    primaryClass = "hep-ph",
    reportNumber = "ACFI-T20-18",
    doi = "10.1103/PhysRevD.103.095027",
    journal = "Phys. Rev. D",
    volume = "103",
    number = "9",
    pages = "095027",
    year = "2021"
}

@article{Joyce:1994bi,
    author = "Joyce, Michael and Prokopec, Tomislav and Turok, Neil",
    title = "{Efficient electroweak baryogenesis from lepton transport}",
    eprint = "hep-ph/9401352",
    archivePrefix = "arXiv",
    reportNumber = "PUPT-1437, PUPT-91-1437",
    doi = "10.1016/0370-2693(94)91377-3",
    journal = "Phys. Lett. B",
    volume = "338",
    pages = "269--275",
    year = "1994"
}

@article{Morrissey:2012db,
    author = "Morrissey, David E. and Ramsey-Musolf, Michael J.",
    title = "{Electroweak baryogenesis}",
    eprint = "1206.2942",
    archivePrefix = "arXiv",
    primaryClass = "hep-ph",
    reportNumber = "NPAC-12-08",
    doi = "10.1088/1367-2630/14/12/125003",
    journal = "New J. Phys.",
    volume = "14",
    pages = "125003",
    year = "2012"
}

@article{Chen:2017bff,
    author = "Chen, Xin and Wu, Yongcheng",
    title = "{Search for CP violation effects in the $h\to \tau\tau$ decay with $e^+e^-$ colliders}",
    eprint = "1703.04855",
    archivePrefix = "arXiv",
    primaryClass = "hep-ph",
    doi = "10.1140/epjc/s10052-017-5258-y",
    journal = "Eur. Phys. J. C",
    volume = "77",
    number = "10",
    pages = "697",
    year = "2017"
}

@article{Chen:2017nxp,
    author = "Chen, Xin and Wu, Yongcheng",
    title = "{Probing the CP-Violation effects in the $h\tau\tau$ coupling at the LHC}",
    eprint = "1708.02882",
    archivePrefix = "arXiv",
    primaryClass = "hep-ph",
    doi = "10.1016/j.physletb.2019.01.038",
    journal = "Phys. Lett. B",
    volume = "790",
    pages = "332--338",
    year = "2019"
}

@article{Jeans:2018anq,
    author = "Jeans, D. and Wilson, G. W.",
    title = "{Measuring the CP state of tau lepton pairs from Higgs decay at the ILC}",
    eprint = "1804.01241",
    archivePrefix = "arXiv",
    primaryClass = "hep-ex",
    reportNumber = "ILD-PHYS-2018-001",
    doi = "10.1103/PhysRevD.98.013007",
    journal = "Phys. Rev. D",
    volume = "98",
    number = "1",
    pages = "013007",
    year = "2018"
}

@article{Altakach:2022ywa,
    author = "Altakach, Mohammad Mahdi and Lamba, Priyanka and Maltoni, Fabio and Mawatari, Kentarou and Sakurai, Kazuki",
    title = "{Quantum information and CP measurement in H\textrightarrow{}\ensuremath{\tau}+\ensuremath{\tau}- at future lepton colliders}",
    eprint = "2211.10513",
    archivePrefix = "arXiv",
    primaryClass = "hep-ph",
    doi = "10.1103/PhysRevD.107.093002",
    journal = "Phys. Rev. D",
    volume = "107",
    number = "9",
    pages = "093002",
    year = "2023"
}

@article{Cohen:1994ss,
    author = "Cohen, Andrew G. and Kaplan, D. B. and Nelson, A. E.",
    title = "{Diffusion enhances spontaneous electroweak baryogenesis}",
    eprint = "hep-ph/9406345",
    archivePrefix = "arXiv",
    reportNumber = "NSF-ITP-94-67, BUHEP-94-11, DOE-ER-40561-146, INT-94-0059, UW-PT-9402",
    doi = "10.1016/0370-2693(94)00935-X",
    journal = "Phys. Lett. B",
    volume = "336",
    pages = "41--47",
    year = "1994"
}

@article{Gritsan:2022php,
    author = "Gritsan, A. V. and others",
    title = "{Snowmass White Paper: Prospects of CP-violation measurements with the Higgs boson at future experiments}",
    eprint = "2205.07715",
    archivePrefix = "arXiv",
    primaryClass = "hep-ex",
    reportNumber = "FERMILAB-PUB-22-429-SCD",
    month = "5",
    year = "2022"
}

@article{Athron:2024rir,
    author = "Athron, Peter and Crivellin, Andreas and Gonzalo, Tom\'as E. and Iguro, Syuhei and Sierra, Cristian",
    title = "{Global fit to the 2HDM with generic sources of flavour violation using GAMBIT}",
    eprint = "2410.10493",
    archivePrefix = "arXiv",
    primaryClass = "hep-ph",
    reportNumber = "gambit-physics-2024",
    doi = "10.1007/JHEP11(2024)133",
    journal = "JHEP",
    volume = "11",
    pages = "133",
    year = "2024"
}

@article{ILCInternationalDevelopmentTeam:2022izu,
    author = "Aryshev, Alexander and others",
    collaboration = "ILC International Development Team",
    title = "{The International Linear Collider: Report to Snowmass 2021}",
    eprint = "2203.07622",
    archivePrefix = "arXiv",
    primaryClass = "physics.acc-ph",
    reportNumber = "DESY-22-045, IFT-UAM/CSIC-22-028, KEK Preprint 2021-61, IFT--UAM/CSIC--22-028, KEK Preprint 2021-61,
  PNNL-SA-160884, SLAC-PUB-17662, FERMILAB-FN-1171-PPD-QIS-SCD-TD, PNNL-SA-160884",
    month = "3",
    year = "2022"
}

@article{FCC:2018evy,
    author = "Abada, A. and others",
    collaboration = "FCC",
    title = "{FCC-ee: The Lepton Collider}: {Future Circular Collider Conceptual Design Report Volume 2}",
    reportNumber = "CERN-ACC-2018-0057",
    doi = "10.1140/epjst/e2019-900045-4",
    journal = "Eur. Phys. J. ST",
    volume = "228",
    number = "2",
    pages = "261--623",
    year = "2019"
}

@article{Roussy:2022cmp,
    author = "Roussy, Tanya S. and others",
    title = "{An improved bound on the electron\textquoteright{}s electric dipole moment}",
    eprint = "2212.11841",
    archivePrefix = "arXiv",
    primaryClass = "physics.atom-ph",
    doi = "10.1126/science.adg4084",
    journal = "Science",
    volume = "381",
    number = "6653",
    pages = "adg4084",
    year = "2023"
}

@article{CEPCStudyGroup:2018ghi,
    author = "Dong, Mingyi and others",
    editor = "Guimar\~aes da Costa, Jo\~ao Barreiro and others",
    collaboration = "CEPC Study Group",
    title = "{CEPC Conceptual Design Report: Volume 2 - Physics \& Detector}",
    eprint = "1811.10545",
    archivePrefix = "arXiv",
    primaryClass = "hep-ex",
    reportNumber = "IHEP-CEPC-DR-2018-02, IHEP-EP-2018-01, IHEP-TH-2018-01",
    month = "11",
    year = "2018"
}

@article{Abel:2020pzs,
    author = "Abel, C. and others",
    title = "{Measurement of the Permanent Electric Dipole Moment of the Neutron}",
    eprint = "2001.11966",
    archivePrefix = "arXiv",
    primaryClass = "hep-ex",
    doi = "10.1103/PhysRevLett.124.081803",
    journal = "Phys. Rev. Lett.",
    volume = "124",
    number = "8",
    pages = "081803",
    year = "2020"
}

@article{Hisano:2012cc,
    author = "Hisano, Junji and Tsumura, Koji and Yang, Masaki J. S.",
    title = "{QCD Corrections to Neutron Electric Dipole Moment from Dimension-six Four-Quark Operators}",
    eprint = "1205.2212",
    archivePrefix = "arXiv",
    primaryClass = "hep-ph",
    reportNumber = "IPMU12-0088",
    doi = "10.1016/j.physletb.2012.06.038",
    journal = "Phys. Lett. B",
    volume = "713",
    pages = "473--480",
    year = "2012"
}

@article{Hisano:2012sc,
    author = "Hisano, Junji and Lee, Jeong Yong and Nagata, Natsumi and Shimizu, Yasuhiro",
    title = "{Reevaluation of Neutron Electric Dipole Moment with QCD Sum Rules}",
    eprint = "1204.2653",
    archivePrefix = "arXiv",
    primaryClass = "hep-ph",
    reportNumber = "IPMU-12-0065, TU-903",
    doi = "10.1103/PhysRevD.85.114044",
    journal = "Phys. Rev. D",
    volume = "85",
    pages = "114044",
    year = "2012"
}

@article{Bertolini:2019out,
    author = "Bertolini, Stefano and Maiezza, Alessio and Nesti, Fabrizio",
    title = "{Kaon CP violation and neutron EDM in the minimal left-right symmetric model}",
    eprint = "1911.09472",
    archivePrefix = "arXiv",
    primaryClass = "hep-ph",
    doi = "10.1103/PhysRevD.101.035036",
    journal = "Phys. Rev. D",
    volume = "101",
    number = "3",
    pages = "035036",
    year = "2020"
}

@article{GAMBIT:2017qxg,
    author = "Bal\'azs, Csaba and others",
    collaboration = "GAMBIT",
    title = "{ColliderBit: a GAMBIT module for the calculation of high-energy collider observables and likelihoods}",
    eprint = "1705.07919",
    archivePrefix = "arXiv",
    primaryClass = "hep-ph",
    reportNumber = "gambit-code-2017",
    doi = "10.1140/epjc/s10052-017-5285-8",
    journal = "Eur. Phys. J. C",
    volume = "77",
    number = "11",
    pages = "795",
    year = "2017"
}

@article{Martinez:2017lzg,
    author = "Martinez, Gregory D. and McKay, James and Farmer, Ben and Scott, Pat and Roebber, Elinore and Putze, Antje and Conrad, Jan",
    collaboration = "GAMBIT",
    title = "{Comparison of statistical sampling methods with ScannerBit, the GAMBIT scanning module}",
    eprint = "1705.07959",
    archivePrefix = "arXiv",
    primaryClass = "hep-ph",
    reportNumber = "gambit-code-2017",
    doi = "10.1140/epjc/s10052-017-5274-y",
    journal = "Eur. Phys. J. C",
    volume = "77",
    number = "11",
    pages = "761",
    year = "2017"
}

@article{GAMBITFlavourWorkgroup:2017dbx,
    author = "Bernlochner, Florian U. and others",
    collaboration = "GAMBIT Flavour Workgroup",
    title = "{FlavBit: A GAMBIT module for computing flavour observables and likelihoods}",
    eprint = "1705.07933",
    archivePrefix = "arXiv",
    primaryClass = "hep-ph",
    reportNumber = "NORDITA-2017-077, CERN-TH-2017-167, NORDITA 2017-077, gambit-code-2017",
    doi = "10.1140/epjc/s10052-017-5157-2",
    journal = "Eur. Phys. J. C",
    volume = "77",
    number = "11",
    pages = "786",
    year = "2017"
}

@article{Engel:2013lsa,
    author = "Engel, Jonathan and Ramsey-Musolf, Michael J. and van Kolck, U.",
    title = "{Electric Dipole Moments of Nucleons, Nuclei, and Atoms: The Standard Model and Beyond}",
    eprint = "1303.2371",
    archivePrefix = "arXiv",
    primaryClass = "nucl-th",
    doi = "10.1016/j.ppnp.2013.03.003",
    journal = "Prog. Part. Nucl. Phys.",
    volume = "71",
    pages = "21--74",
    year = "2013"
}

@article{Goncalves:2021egx,
    author = "Gon\c{c}alves, Dorival and Kaladharan, Ajay and Wu, Yongcheng",
    title = "{Electroweak phase transition in the 2HDM: Collider and gravitational wave complementarity}",
    eprint = "2108.05356",
    archivePrefix = "arXiv",
    primaryClass = "hep-ph",
    doi = "10.1103/PhysRevD.105.095041",
    journal = "Phys. Rev. D",
    volume = "105",
    number = "9",
    pages = "095041",
    year = "2022"
}

@article{Biekotter:2022kgf,
    author = {Biek\"otter, Thomas and Heinemeyer, Sven and No, Jos\'e Miguel and Olea-Romacho, Mar\'\i{}a Olalla and Weiglein, Georg},
    title = "{The trap in the early Universe: impact on the interplay between gravitational waves and LHC physics in the 2HDM}",
    eprint = "2208.14466",
    archivePrefix = "arXiv",
    primaryClass = "hep-ph",
    reportNumber = "DESY-22-127, IFT--UAM/CSIC--22-015",
    doi = "10.1088/1475-7516/2023/03/031",
    journal = "JCAP",
    volume = "03",
    pages = "031",
    year = "2023"
}

@article{Su:2020pjw,
    author = "Su, Wei and Williams, Anthony G. and Zhang, Mengchao",
    title = "{Strong first order electroweak phase transition in 2HDM confronting future Z \& Higgs factories}",
    eprint = "2011.04540",
    archivePrefix = "arXiv",
    primaryClass = "hep-ph",
    reportNumber = "ADP-20-31/T1141",
    doi = "10.1007/JHEP04(2021)219",
    journal = "JHEP",
    volume = "04",
    pages = "219",
    year = "2021"
}

@article{Dorsch:2013wja,
    author = "Dorsch, G. C. and Huber, S. J. and No, J. M.",
    title = "{A strong electroweak phase transition in the 2HDM after LHC8}",
    eprint = "1305.6610",
    archivePrefix = "arXiv",
    primaryClass = "hep-ph",
    doi = "10.1007/JHEP10(2013)029",
    journal = "JHEP",
    volume = "10",
    pages = "029",
    year = "2013"
}

@article{Basler:2016obg,
    author = "Basler, P. and Krause, M. and Muhlleitner, M. and Wittbrodt, J. and Wlotzka, A.",
    title = "{Strong First Order Electroweak Phase Transition in the CP-Conserving 2HDM Revisited}",
    eprint = "1612.04086",
    archivePrefix = "arXiv",
    primaryClass = "hep-ph",
    doi = "10.1007/JHEP02(2017)121",
    journal = "JHEP",
    volume = "02",
    pages = "121",
    year = "2017"
}

@article{Bernon:2017jgv,
    author = "Bernon, J\'er\'emy and Bian, Ligong and Jiang, Yun",
    title = "{A new insight into the phase transition in the early Universe with two Higgs doublets}",
    eprint = "1712.08430",
    archivePrefix = "arXiv",
    primaryClass = "hep-ph",
    doi = "10.1007/JHEP05(2018)151",
    journal = "JHEP",
    volume = "05",
    pages = "151",
    year = "2018"
}

@article{Dorsch:2017nza,
    author = "Dorsch, G. C. and Huber, S. J. and Mimasu, K. and No, J. M.",
    title = "{The Higgs Vacuum Uplifted: Revisiting the Electroweak Phase Transition with a Second Higgs Doublet}",
    eprint = "1705.09186",
    archivePrefix = "arXiv",
    primaryClass = "hep-ph",
    reportNumber = "CP3-17-15, DESY-17-076, KCL-PH-TH-2017-27",
    doi = "10.1007/JHEP12(2017)086",
    journal = "JHEP",
    volume = "12",
    pages = "086",
    year = "2017"
}

@article{Andersen:2017ika,
    author = "Andersen, Jens O. and Gorda, Tyler and Helset, Andreas and Niemi, Lauri and Tenkanen, Tuomas V. I. and Tranberg, Anders and Vuorinen, Aleksi and Weir, David J.",
    title = "{Nonperturbative Analysis of the Electroweak Phase Transition in the Two Higgs Doublet Model}",
    eprint = "1711.09849",
    archivePrefix = "arXiv",
    primaryClass = "hep-ph",
    reportNumber = "HIP-2017-26/TH, HIP-2017-26-TH",
    doi = "10.1103/PhysRevLett.121.191802",
    journal = "Phys. Rev. Lett.",
    volume = "121",
    number = "19",
    pages = "191802",
    year = "2018"
}
\end{document}